\documentclass[lettersize,journal]{IEEEtran}
\usepackage{amsmath,amsfonts}
\usepackage{array}
\usepackage{textcomp}
\usepackage{url}
\usepackage{verbatim}
\usepackage{graphicx}
\usepackage{cite}
\usepackage{amsmath}
\usepackage{multirow}
\usepackage{amsfonts,amsmath}
\usepackage{amssymb, nccmath}
\usepackage{braket}
\usepackage{amssymb, nccmath}
\usepackage{dsfont}
\usepackage{booktabs}
\usepackage[utf8]{inputenc}
\usepackage{dsfont}
\usepackage{amsfonts}
\usepackage{booktabs}
\usepackage{graphicx}
\usepackage{grffile}
\usepackage{grffile}
\usepackage{array} 
\usepackage{mathtools}
\usepackage{subfigure} 
\usepackage{fontenc}
\usepackage{tikz}
\usepackage{pgfplots}
\usepackage{tikz}
\newtheorem{definition}{Definition}
\DeclareMathAlphabet\mathchorus     {T1}{qzc} {m} {n}
\usetikzlibrary{patterns}
\definecolor{myhund}{HTML}{9966CC}
\definecolor{myfifty}{HTML}{002E63}
\definecolor{mytwenty}{HTML}{4997D0}
\definecolor{s1}{HTML}{66FF00}
\definecolor{s2}{HTML}{00CC99}
\definecolor{s3}{HTML}{8DB600}
\definecolor{s4}{HTML}{177245}
\definecolor{r1}{HTML}{FF004F}
\definecolor{r2}{HTML}{FF4F00 }
\definecolor{r3}{HTML}{F400A1}
\definecolor{r4}{HTML}{CE2029}
\usetikzlibrary{positioning}
\definecolor{Mycolor}{HTML}{03C03C}
\usepackage[linesnumbered, ruled]{algorithm2e}
\makeatletter
\newcommand{\removelatexerror}{\let\@latex@error\@gobble}
\def\ps@IEEEtitlepagestyle{%
	\def\@oddfoot{\mycopyrightnotice}%
	\def\@oddhead{\hbox{}\@IEEEheaderstyle\leftmark\hfil\thepage}\relax
	\def\@evenhead{\@IEEEheaderstyle\thepage\hfil\leftmark\hbox{}}\relax
	\def\@evenfoot{}%
}

\def\mycopyrightnotice{%
	\begin{minipage}{\textwidth}
		\centering \scriptsize
		 © 2025 IEEE. This article has been accepted in IEEE Transactions on Cloud Computing Journal © 2025 IEEE. Personal use of this material is permitted. Permission from IEEE must be obtained for all other uses, in any current or future media, including reprinting/republishing this material for advertising or promotional purposes, creating new collective works, for resale or redistribution to servers or lists, or reuse of any copyrighted component of this work in other works. This work is freely available for survey and citation.
	\end{minipage}
}
\makeatother
\begin{document}

\title{REE-TM: Reliable and Energy-Efficient Traffic Management Model for Diverse Cloud Workloads}
\author{Ashutosh~Kumar~Singh,~\textit{Senior Member, IEEE},~Deepika~Saxena,~\textit{Member, IEEE},~Volker~Lindenstruth
	\IEEEcompsocitemizethanks{\IEEEcompsocthanksitem 
 A. K. Singh is with Indian Institute of Information Technology Bhopal, India and VIZJA University, Warsaw, Poland, Europe. E-mail:  director@iiitbhopal.ac.in \\
D. Saxena is with the School of Computer Science and Engineering, The University of Aizu, Japan and VIZJA University, Warsaw, Poland, Europe. E-mail: 13deepikasaxena@gmail.com \\
		V. Lindenstruth is with the Department of Computer Science and Engineering, Goethe University, Frankfurt, Germany, E-mail: voli@compeng.de 
}}



\maketitle

\begin{abstract}
	Diversity of workload demands lays a critical impact on efficient resource allocation and management of cloud services. The existing literature has either weakly considered or overlooked the heterogeneous feature of job requests received from wide range of internet services users. To address this context, the proposed  approach named \textbf{R}eliable and \textbf{E}nergy \textbf{E}fficient \textbf{T}raffic \textbf{M}anagement (\textbf{REE-TM}) has exploited the diversity of internet traffic in terms of variation in resource demands and expected complexity. {Specifically, REE-TM incorporates categorization of heterogeneous job requests and executes them by selecting the most admissible \textit{virtual node} (a software-defined instance such as a virtual machine or container) and \textit{physical node} (an actual hardware server or compute host) within the cloud infrastructure.} To deal with resource-contention-based resource failures and performance degradation, a novel workload estimator `Toffoli Gate-based Quantum Neural Network' (TG-QNN) is proposed, wherein learning process or interconnection weights optimization is achieved using Quantum version of BlackHole (QBHO) algorithm. The proactively estimated workload is used to compute entropy of the upcoming internet traffic with various traffic states analysis for detection of probable resource-congestion. REE-TM is extensively evaluated through simulations using a benchmark dataset and compared with optimal and without REE-TM versions. The performance evaluation and comparison of REE-TM with measured significant metrics reveal its effectiveness in assuring higher reliability by up to 30.25\% and energy-efficiency by up to 23\% as compared without REE-TM.
\end{abstract}


\section{Introduction}
\IEEEPARstart{C}{loud}-based heterogeneous services, encompassing IoT, cyber-physical systems, and edge computing, have transformed internet services across sectors like corporate, social networks, and scientific research \cite{chakraborty2024optimizing}. These technologies enable diverse devices, such as smart IoT applications and industrial machines, to seamlessly communicate and exchange information \cite{kaur2021energy}. Equipped with embedded sensors, these devices capture real-world data (text, audio, multimedia) and respond intelligently \cite{al2019energy,kang2021adaptive}. Integrating cloud computing with big data analytics and predictive modeling optimizes performance and enhances service delivery. In IoT infrastructures, data from smart devices is processed through cloud nodes to generate insights and support decision-making. This interconnected system, known as a `\textit{Diverse Cloud Services Environment}', leverages cloud computing for efficient data storage, processing, and transmission. Cloud services, delivered via internet-connected servers, establish standards for connecting, monitoring, and securing devices. However, efficient real-time data processing in cloud centers faces challenges due to limited physical resource capacity and dynamic fluctuations in workloads from various sources. The diverse internet traffic, with varying resource demands, priorities, and execution times, necessitates dynamic management to minimize delays and ensure high service quality \cite{gai2017sa}. The major challenges due to the diversity of cloud workload demands include:

\begin{itemize}
    \item \textit{Varied resource requirements}: Diverse workloads require different levels of compute, storage, and network resources, complicating traffic optimization without over-provisioning or under-utilization.
    \item \textit{Dynamic Traffic Patterns}: Workload diversity induces fluctuating traffic, demanding adaptive strategies to manage sudden spikes or drops without sacrificing performance or energy efficiency.
    \item \textit{Latency sensitivity}: Real-time applications are latency sensitive, necessitating precise traffic routing and prioritization to meet stringent performance demands.
    \item \textit{Complex management}: Maintaining consistent Quality of Service (QoS) across diverse workloads is challenging due to varying priorities, requiring intelligent traffic management to balance competing demands.
\end{itemize}
{
A recent 2024 survey on cloud service outages~\cite{cloudoutages2024} reported the year's largest disruption, where Microsoft Azure faced a 72-hour downtime, affecting global services and resulting in an estimated loss of \$10 billion. High reliability is critical for cloud services that support diverse applications that store and transmit data in formats such as images, audio, video, and animations. However, factors such as as machine failures, network congestion, resource contention, and mismanagement can severely affect service continuity. Improving the availability and reliability of cloud infrastructure is not only vital for performance but also directly influences energy efficiency. Unreliable systems typically require more energy to recover from failures, rerun processes, or address inefficiencies caused by over-provisioning resources to compensate for these faults. In addition, unreliable service delivery leads to more frequent error handling, redundant task executions, and longer recovery times, all of which contribute to increased energy consumption. In contrast, a reliable cloud environment minimizes downtime, optimizes resource usage, and ensures energy-efficient task scheduling and execution. In essence, improving system reliability helps reduce unnecessary energy consumption by preventing idle resources, minimizing system overhead, and reducing the need for backup systems or manual interventions.}

{For cloud services handling real-time data from industrial systems, IoT devices, and critical applications in sectors such as healthcare and defense, this relationship between reliability and energy efficiency becomes even more crucial. The varying service times, resource needs, and deadlines for such tasks necessitate energy-efficient strategies that do not compromise reliability.} The main challenges include:
\begin{itemize}
    \item \textit{Optimal resource allocation}: Efficient resource management ensures energy savings by avoiding resource over-provisioning and under-utilization while meeting diverse application demands.
    \item \textit{Real-time data processing}: Low-latency and energy-efficient processing are key to maintaining service reliability while minimizing energy costs for high-throughput data streams.
    \item \textit{Continuous service availability}: Reliable fault-tolerance mechanisms and rapid recovery strategies help maintain service availability without excessive energy expenditure, especially during failure scenarios.
\end{itemize}
Therefore, addressing these challenges is essential for creating cloud infrastructures that are both reliable and energy-efficient.

\subsection{Related work}
Several approaches tackled resource management in the cloud computing environments.These approaches include \textit{proactive methods}, such as workload balancing, rely on prior knowledge of workload and applications \cite{saxena2021op, saxena2022high,saxena2022fault,gai2017sa}, and \textit{reactive approaches}, which handles traffic load at actual arrival by resource allocation and job scheduling \cite{kaur2021energy,kaur2019keids}.  Saxena et al. \cite{saxena2022fault} proposed a Fault Tolerant Elastic Resource Management (FT-ERM) approach, integrating failure tolerance into workload distribution among physical and virtual nodes, while Al et al. \cite{al2019energy} presented optimization scenarios for reliable IoT environments, including standby routes selection and reliability-based data compression. Ming et al. \cite{ming2023adaptive} introduced a general multitasking framework for solving constrained multi-objective optimization problems such as cloud resource allocation using reinforcement learning to select the most suitable auxiliary tasks during the evolutionary process. 
Saxena et al. \cite{saxena2022high} introduced a VM Significance Ranking Estimation-based High Availability Management (SRE-HM) model for maximizing internet service availability, considering VM communication frequency. Gai et al. \cite{gai2017resource} focused on sustainability, utilizing energy-efficient cloud computing in their Smart Cloud-based Optimizing Workload (SCOW) model for IoT application processing. Kaur et al. \cite{kaur2021energy} proposed an energy-efficient cloud-based cyberphysical system resource management framework, integrating multi-objective job scheduling and energy-aware scheduling. Nesmachnow et al. \cite{nesmachnow2015efficient} proposed a cloud broker that outsources reserved VMs to customers at lower prices than cloud providers, leveraging the price difference between on-demand and reserved VMs. It developed a smart heuristics to maximize broker profit, with experimental evaluations demonstrating effectiveness using real-world data.
Saxena et al. \cite{saxena2021op} presented an Online Prediction-based Multi-objective Load Balancing (OP-MLB) framework for resource utilization improvement, utilizing evolutionary neural networks for workload prediction. Additionally, Kaur et al. \cite{kaur2019keids} introduced KEIDS, a Kubernetes container management solution addressing interference and energy minimization in industrial IoT setups, while Manogaran et al. \cite{manogaran2022optimal} proposed a decisive energy management scheme for optimal resource allocation in IoT environments. Muthanna et al. \cite{muthanna2019secure} introduced an IoT traffic management framework deploying edge computing and Software-Defined Network (SDN) for effective reliability. \\
{ A novel cloud workload prediction model TFEGRU is proposed in \cite{zhao2024tfegru} that integrates a Time-Frequency Enhanced Block  with GRU and multi-head attention to capture complex temporal-frequency patterns. It addressed high-dimensional workload challenges using channel-independent strategies and embedding to enhance predictive accuracy. Li et al. \cite{li2024evogwp} proposed EvoGWP which is a graph-based evolution learning approach for long-term cloud workload prediction that captures dynamic usage pattern changes. It leverages a spatio-temporal graph neural network (GNN) with shapelet extraction to model both temporal trends and spatial interference across workloads. Recently, quantum-inspired approaches have emerged as powerful tools for workload forecasting in cloud environments. For instance, Singh et al.~\cite{singh2021quantum} presented an Evolutionary Quantum Neural Network (EQNN), leveraging qubit diversity and quantum operations to boost forecasting performance. Building on these advancements, Gupta et al.~\cite{gupta2024multiple} introduced the Multiple Controlled Toffoli-driven Adaptive Quantum Neural Network (MCT-AQNN), which effectively models temporal workload variations for more precise predictions.
}

\subsection{Observed Research Gaps}
Despite extensive research, a significant gap remains in addressing the inherent diversity and variability in cloud traffic workloads, which is vital for ensuring reliable and efficient workload management. Current solutions often rely on static or generalized assumptions about workload characteristics, which fail to capture the complex, dynamic nature of modern cloud environments. For example, traditional models may assume homogeneity in resource demands, leading to suboptimal resource allocation, increased latency, and potential bottlenecks during peak traffic periods \cite{al2019energy}. Furthermore, many existing methods lack adaptive mechanisms to cope with sudden surges in workload diversity, resulting in compromised service reliability and increased energy consumption \cite{gai2017sa}. The increasing integration of IoT, edge computing, and cyber-physical systems into cloud infrastructures exacerbates these challenges, as these systems generate highly variable and often unpredictable traffic patterns \cite{kaur2021energy}. Moreover, the failure to consider workload heterogeneity can lead to inefficient use of resources, particularly in environments where different workloads exhibit vastly different resource consumption profiles \cite{manogaran2022optimal}. This inefficiency can manifest in over-provisioning, under-utilization, and higher operational costs, ultimately impacting the overall performance and reliability of cloud services \cite{nesmachnow2015efficient}.
\subsection{Our Contributions}
To address above mentioned challenges, this paper proposes multifaceted \textbf{R}eliable and \textbf{E}nergy-\textbf{E}fficient \textbf{T}raffic \textbf{M}anagement (\textbf{REE-TM}) approach, which introduces a novel method for managing heterogeneous cloud traffic with a focus on maximizing reliability and resource efficiency. Unlike prior methods, REE-TM employs a Quantum Machine Learning-based Workload Estimator (QML-WE) to dynamically assess resource capacity usage across a wide spectrum of cloud workloads. QML-WE leverages quantum computing principles to handle the high-dimensional, complex data characteristic of heterogeneous cloud environments, providing more accurate predictions of resource needs under varying traffic conditions. Additionally, the proposed model incorporates a diverse Traffic States Computation-assisted Entropy Analyser (TSC-EA), which quantifies the randomness and variability in internet traffic states. This entropy-based analysis enables the model to detect anomalies and adaptively reallocate resources in real time, ensuring optimal performance even under highly fluctuating traffic loads \cite{manogaran2022optimal}. By differentiating traffic states based on anticipated resource usage, REE-TM enhances resource planning, improves service reliability, and optimizes energy consumption, making it particularly suitable for environments where consistent performance and high availability are critical. The key contributions of the proposed model are fourfold:

\begin{itemize}
	\item To procure reliability and sustainability of cloud traffic networks, a  Toffoli gate embedded  Quantum Neural Network (TG-QNN) based workload analyser is proposed in which the learning process is achieved by newly developed Quantum Blackhole Optimization (QBHO) algorithm. It helps to analyse the resource capacity usage of heterogeneous internet traffic and  detect the probable congestion for approaching traffic and alleviate its worse effect by employing admissible management decisions.  
	
	\item A traffic states computation engine is developed that examines the entropy of the internet traffic in real-time and seggregates it into three distinct traffic states namely, Stragglers, Resource Hogs, and Normal traffic. This functionality enables efficient and admissible handling of the miscellaneous internet traffic.
	
	\item Resource capacity-sensitive jobs demarcation suited for the modeling of reliability and sustainability is introduced for heterogeneous traffic management to  improve the quality of service experience for the end users.

	\item Implementation and evaluation of proposed approach by using real benchmark dataset reveal that REE-TM outperforms the state-of-the-art approaches in terms of various performance metrics.  
	
\end{itemize}

\begin{figure}[!htbp]
	\centering
	\includegraphics[width=0.9\linewidth]{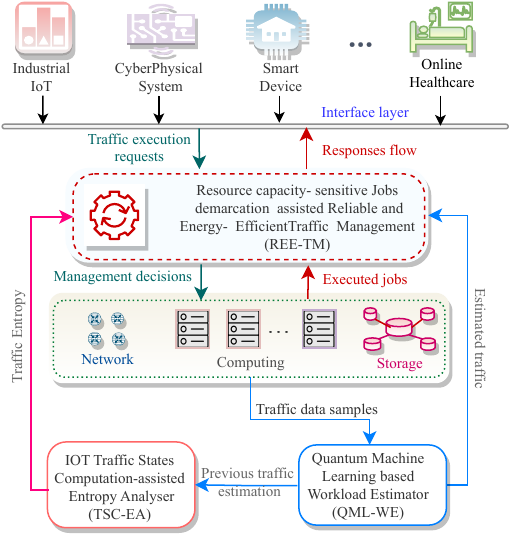}
	\caption{Schematic design of REE-TM model}
	\label{fig:birdeyeview}
\end{figure}

The concise schematic design of the REE-TM model, illustrated in Fig. \ref{fig:birdeyeview},  where extensive range of internet workloads, including industrial IoT, cyberphysical systems, smart devices, and online healthcare services, submitting their application execution requests through an interface layer to the cloud infrastructure. Resource capacity-sensitive jobs are segregated to enable effective resource capacity planning and management. Within the cloud computing environment, comprising network, computing, and storage components, internet traffic applications are executed, and resource usage data samples are collected for traffic estimation by the QML-WE unit in the subsequent interval. TSC-EA leverages previous traffic estimation values to compute the entropy of upcoming traffic and categorize it into different traffic states.

\subsection{Motivation for TG-QNN}

The conventional machine learning methods, used in previous research \cite{kaur2019keids,kaur2021energy,  saxena2022high, saxena2022fault, saxena2021op}, are inadequate to fully capture and leverage the complex correlations among patterns, limiting their accuracy in predicting and analyzing heterogeneous internet traffic within high-dimensional and variable cloud workloads. This challenge motivates the development of a Quantum Toffoli Gate (QTG)-embedded Neural Network, which excels in handling intricate relationships through its ability to perform complex conditional operations using two control qubits. The QTG's three-qubit interaction capability provides superior management of data correlations, making it more effective than the Quantum C-NOT gate \cite{singh2021quantum} in modeling dynamic and fluctuating workloads. Furthermore, the integration of Quantum Blackhole optimization into the QTG neural network enhances learning efficiency and speeds up convergence, optimizing performance in dynamic data environments. This innovative approach ensures accurate estimation of workload ranges and effective reservation of cloud resources, addressing the limitations of conventional methods and providing a robust solution for managing heterogeneous cloud traffic.
\subsection{Paper Organization}
The rest of the paper is structured as follows: Section II provides an overview of the proposed REE-TM approach, which includes three main units: Cloud Traffic Management, Workload Estimation, and Traffic States Computation-Assisted Entropy Analysis. These units are detailed in Sections III, IV, and V, respectively. Section VI discusses the operational design and complexity of REE-TM. Performance evaluation and comparative analysis are presented in Section VII, followed by the conclusion in Section VIII. {
The symbols alongwith their meaning used throughout the paper are described in Table \ref{tab:notation}}.

{
\begin{table}[!htbp]
\centering
\caption{{List of Symbols and Notations}}
\label{tab:notation}
  \resizebox{0.5\textwidth}{!}{
\begin{tabular}{lrlr}
\hline
        Symbol& Meaning & Symbol & Meaning \\ \hline
        $m$ & number of jobs & $\lambda$ & request  \\
        $\Omega$ & response &  $\mathchorus{FT}$ & functional tasks \\
        $\mathchorus{D}$ & data sample & $\mathds{JR}$ & job request \\
        
        $St$ & straggler & $Nt$ & normal tasks\\
        $Rh$ & resource hog & $\mathds{C}$ & cluster of tasks \\ $\mathchorus{C}$,$\mathchorus{M}$,$\mathchorus{B}$ & CPU, memory, bandwidth & $R$ & resource \\
        $\mathchorus{VN}$ & virtual node & $\mathchorus{PN}$  & physical node\\
        
        $\omega$ & allocation of $Q$  $\mathchorus{VN}$ among $P$ $\mathchorus{PN}$ & $\mathds{R}$ & global reliability \\

        $\mathchorus{RU}$ & resource utilization  &  $\mathchorus{H}z$ &hazard or failure rate \\
       $\Phi_i$ & $i^{th} qubit neuron$  &  $\Theta$ & qubit weights \\
       
        $\varpi$ & qubit summation   &  $\mathchorus{SIG}()$ &activation function  \\

       $QC_{k}^{best}$ & local blackhole   &  $QG^{best}$ & global best \\
       \hline
\end{tabular}}
\end{table}
}
   
\section{REE-TM Approach}
\begin{figure*}[!htbp]
	\centering
	\includegraphics[width=0.74\linewidth]{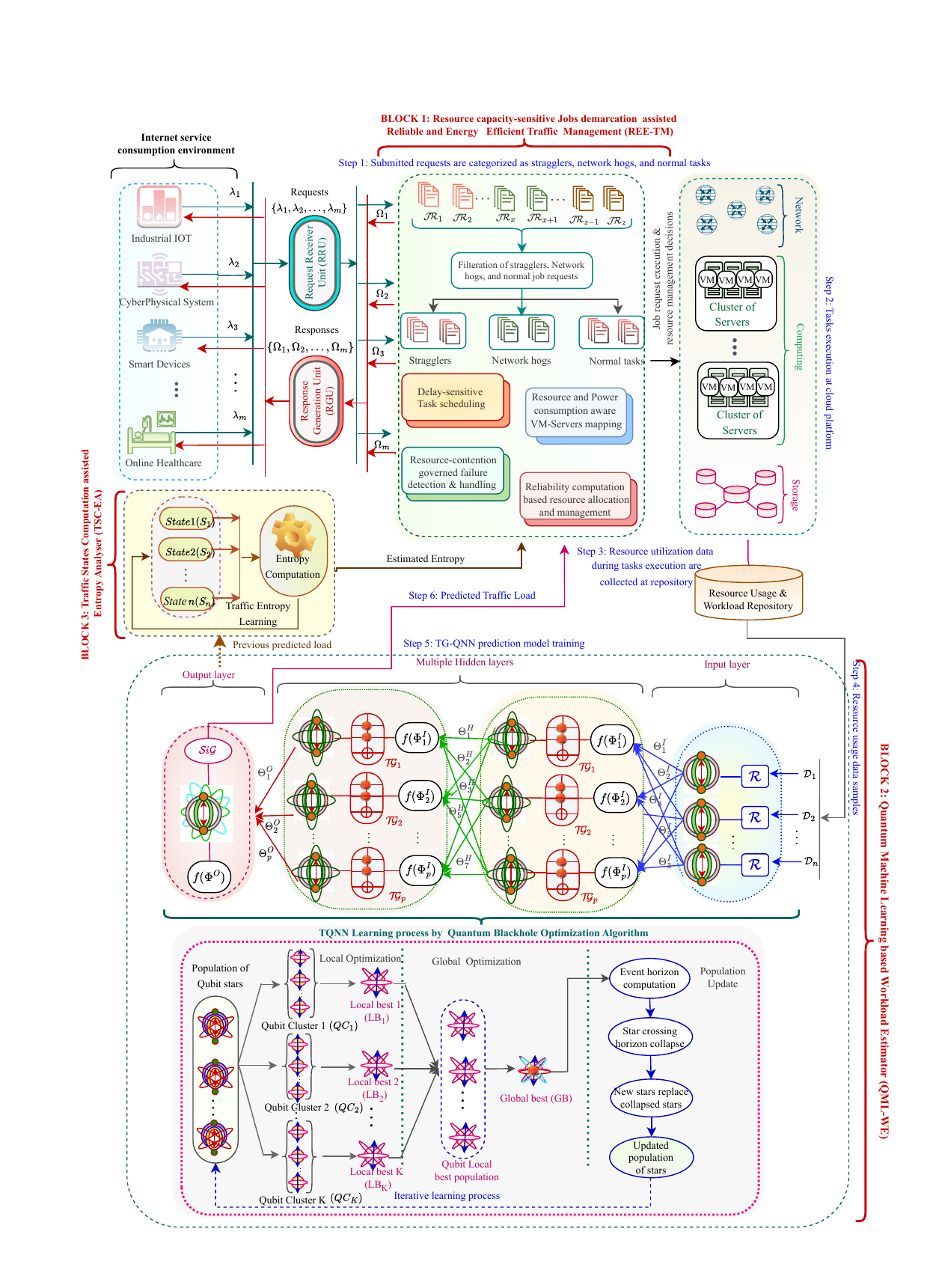}
	\caption{REE-TM Operational Design}
	\label{fig:proposed-model}
\end{figure*}

The complete design and operational flow of the proposed heterogeneous cloud traffic management approach with information flow among the involved entities and block units (viz., BLOCKS 1-3) is depicted in Fig. \ref{fig:proposed-model}.
Consider $m$ job requests \{$\lambda_1$, $\lambda_2$, ..., $\lambda_m$\} are received by the \textit{Request Receiver Unit} (RRU) from heterogeneous internet service consumption infrastructure including Industrial IoT, Cyberphysical system, Smart devices, Online health care services, etc. for execution within the cloud environment. The responses \{$\Omega_1$, $\Omega_2$, ..., $\Omega_m$\} are generated by the \textit{Response Generation Unit} (RGU) after complete processing of the received job requests and passed to the internet service consumption environment. A \textit{Reliable and Energy Efficient Traffic Management} (REE-TM) unit (BLOCK 1) is employed for managing the  cloud traffic by  distributing the heterogeneous job requests \{$\lambda_1$, $\lambda_2$, ..., $\lambda_m$\} for execution in the form of functional tasks \{$\mathchorus{FT}_1$, $\mathchorus{FT}_2$, ..., $\mathchorus{FT}_z$\} (where $z$ is the total number of functional tasks) on various virtual nodes including virtual machines and containers.  REE-TM analyses resource (viz., CPU, memory, bandwidth) requirements of  these funtional tasks  and allows segregation of the resource capacity-sensitive jobs by grouping these functional tasks  into clusters as per their estimated resource usage. Specifically, the functional tasks are assorted into three categories: \textit{Stragglers}, \textit{Resource hogs}, and \textit{Normal jobs}, which are defined as follows:

{
\begin{definition}
\textbf{Stragglers}: Tasks that execute significantly slower than peer tasks despite similar resource allocations, causing delays in distributed cloud workloads are known as \textit{stragglers}.
\end{definition}
\begin{definition}
\textbf{Resource hogs}: The tasks which consume disproportionately high amounts of compute, memory, storage, or network resources, leading to contention and degraded system performance are defined as \textit{resource hogs}.
\end{definition}
\begin{definition}
\textbf{Normal jobs}: The tasks that consume resources within expected thresholds and complete execution predictably without causing contention or delays are \textit{normal jobs}.
\end{definition}
}

These tasks are assigned to various virtual nodes accordingly to enable effective management of cloud physical resources (viz., computing, network, storage) and improve usage of physical resources. Further, REE-TM includes provision for  traffic classification-based task scheduling, resource and power consumption aware virtual to physical nodes mapping, resource-contention governed failure detection and handling,  and  reliability computation based management during physical resource allocation.  The detailed description of REE-TM is entailed in Section \ref{reetm}. 
\par    
The proactive estimation of the approaching network traffic load and the detection of potential job failures due to resource contention are performed by the \textit{Quantum Machine Learning-based Workload Estimation} (QML-WE) unit (Fig.~\ref{fig:proposed-model}, Block 2). A Toffoli-Gate ($\mathchorus{TG}$) embedded Quantum Neural Network (TG-QNN) is proposed, comprising $n$ input qubit neurons, multiple hidden layers with $p$ qubit neurons each, and a single output layer with $q$ qubit neurons. The inter-layer connection weights are represented as qubits and are dynamically adjusted using Quantum Toffoli Gates embedded within the hidden and output layers. To train TG-QNN, a \textit{Quantum version of the BlackHole Optimization} (QBHO) algorithm is developed, comprising three improved stages: \textit{Local Optimization}, \textit{Global Optimization}, and \textit{Population Update}. This enhancement addresses premature convergence issues found in the classical BlackHole Optimization algorithm, which traditionally includes only population optimization and update phases.

{Following the functional task scheduling and resource-sensitive allocation in Block 1, the system monitors the execution of each task, collecting live resource usage samples,  specifically CPU utilization, memory consumption, and network bandwidth usage. These metrics are normalized and aggregated over fixed intervals, forming a sequence of resource usage snapshots ${\mathchorus{D}_1, \mathchorus{D}_2, \mathchorus{D}_3, \ldots, \mathchorus{D}_N}$, where each $\mathchorus{D}_i$ represents the operational state within a specific time window. Before input into TG-QNN, each $\mathchorus{D}_i$ is transformed into a qubit vector $\Phi_i$ using a quantum rotation gate, defined as $\Phi_i = \frac{\pi}{2} \times \mathchorus{D}i$. This quantum representation enables TG-QNN to learn the evolving workload dynamics and proactively forecast the future resource usage trend, specifically predicting the ${N+1}^{\text{th}}$ sample (${\mathchorus{D}_{N+1}}^{{Pr}}$). The predicted trends generated by Block 2 are fed back to Block 1 to enable dynamic cloud traffic management. Based on these forecasts, Block 1 proactively distributes heterogeneous incoming job requests ${\lambda_1, \lambda_2, \ldots, \lambda_m}$ into functional tasks across various virtual nodes (e.g., virtual machines and containers). This feedback-driven orchestration facilitates task clustering (e.g., Stragglers, Resource Hogs, and Normal Jobs) and resource capacity-aware scheduling, thereby improving the overall utilization efficiency of compute, network, and storage resources.}
\par 
A \textit{Traffic States Entropy  Computation Engine} (TSECE) (depicted in Fig. \ref{fig:proposed-model}, BLOCK 3) unit is developed and utilized to control and minimize the worse-effects of traffic load prediction errors during physical resource management in real-time.  
This unit has the ability to compute and learn the entropy (i.e., sudden variation, peak, and downfall) of cloud traffic periodically and suggests adaptation of precautionary operations endeavor to achieve an efficient resource utilization and enhanced QoS. TSECE enables  to learn the entropy of the live traffic  for  analysis of the entropy or fluctuation of anticipated traffic which helps to upgrade the load management, minimizing power and resource wastage by admitting the required and sufficient number of physical nodes in active mode while maintaining QoS. The detailed function of TSECE is given in  Section \ref{tscea}.

\section{ Traffic Management} \label{reetm}
Let the  internet service consumption environments have sent heterogeneous job requests $\mathds{JR} =$ \{$\lambda_1$, $\lambda_2$, ..., $\lambda_m$\} to REE-TM unit for execution into the cloud environment. REE-TM demarcates these job requests $\mathds{JR}$ into distinct categories: \textit{Stragglers} ($St$), \textit{Resource hogs} ($Rh$), and \textit{Normal tasks} ($Nt$) on the basis of their physical resource requirement. Assume $\lambda_k^{\mathchorus{C}}$, $\lambda_k^{\mathchorus{M}}$, and $\lambda_k^{\mathchorus{B}}$ represents CPU, memory, and bandwidth requirements, respectively of the $k^{th}$ request, such that  \{$\lambda_1^{\mathchorus{C}}$, $\lambda_1^{\mathchorus{M}}$, $\lambda_1^{\mathchorus{B}}$\}, \{$\lambda_2^{\mathchorus{C}}$, $\lambda_2^{\mathchorus{M}}$, $\lambda_2^{\mathchorus{B}}$\}, ..., \{$\lambda_m^{\mathchorus{C}}$, $\lambda_m^{\mathchorus{M}}$, $\lambda_m^{\mathchorus{B}}$\}   be the set of the required  resource usage capacity of $m$ requests to be satisfied by the $Q$ virtual nodes \{$\mathchorus{VN}_1$, $\mathchorus{VN}_2$, ..., $\mathchorus{VN}_Q$\} for execution. REE-TM utilizes a classification mechanism such that the jobs with high memory requirement or consuming more time in input/output  than processing are confined to cluster of stragglers ($\mathds{C}^{St}$); likewise,  all the job requests with comparatively higher resource capacity requirement belongs to resource hog cluster ($\mathds{C}^{Rh}$);  and remaining job requests are placed in cluster of normal job requests ($\mathds{C}^{Nt}$). Eq. (\ref{trafficstates}) formulates the categorization  and identification of various traffic states (${\mathds{C}_i^{S}}$) for each job request ($\lambda_k:\forall_k \in \{1,2, ..., m\}$), wherein $\lambda_k^{{R_i}}$ and $\lambda_k^{I/O}$ are the resource requirement (of $i^{th}$ resource viz., CPU, memory, and bandwidth) and input/output or data communication overhead, respectively of the $k^{th}$ request ($\lambda_k$); $\mathchorus{C}^{thr}$, $\mathchorus{M}^{thr}$, and $\mathchorus{B}^{thr}$ represent threshold values for  resource ($R$) viz., CPU, memory, and bandwidth, respectively.  
\begin{multline}\label{trafficstates}
{\mathds{C}_i^{S}}=	\begin{cases}
\mathds{C}^{St} (1), & {\text{If}(\lambda_k^{{R_i}} < R_i^{thr}\quad\&\&\quad \lambda_k^{I/O} \geq {I/O}_k^{thr})} \\
{\mathds{C}^{Rh}} (2), & {\text{If}(\lambda_k^{{\mathchorus{C}}} \geq \mathchorus{C}^{thr} || \lambda_m^{{\mathchorus{M}}} \geq \mathchorus{M}^{thr} || \lambda_m^{{\mathchorus{B}}} \geq \mathchorus{B}^{thr})} \\
\mathds{C}^{Nt} (3) & {\text{Otherwise}}   
\end{cases}  
\\ \forall_k \in \{1,2, ..., m\}; R_i \in \{\mathchorus{C}, \mathchorus{M}, \mathchorus{B}\} 
\end{multline}
The execution of job requests belonging to different traffic states is distinctly managed as per the required quality of service  of the associated environment. Further, the common critical concerns of different services environments such as reliability and energy-efficiency are incorporated during virtual nodes allocations and physical nodes management. The process of request execution is discussed in the following subsection \ref{Request Execution}, which is consecutively preceded by reliability modeling (subsection \ref {rem}) and energy-efficiency modeling (subsection \ref{eem}).

\subsection{Request Assignment and Execution} \label{Request Execution}
The  assignment of $k^{th}$ job request to $j^{th}$ virtual node ($\mathchorus{VN}_j$) hosted on $i^{th}$ physical node ($\mathchorus{PN}_i$) is represented as mapping ($\omega_{kji}$) which is mathematically formulated using Eqs. (\ref{re1}-\ref{re3}) subject to constraints ($C_1-C_6$) as mentioned in Eq. (\ref{cs}). Accordingly,  the job requests confined to cluster of stragglers ($\mathds{C}_{St}$) are assigned to virtual nodes hosted on a physical node with sufficient CPU as well as memory capacity and allows needed I/O operations ($MAX(List_{\mathchorus{PN}_i^{\mathchorus{B}}})$). While the resource hogs ($\mathds{C}_{Rh}$) are allocated to physical nodes having  higher resource capacity ($R$) comparatively ($MAX(List_{\mathchorus{PN}_i^{R} })$) that  can serve the requirement. All the remaining job requests are considered as `normal'  to be scheduled on the remaining physical nodes having higher processing speed ($MAX(List_{\mathchorus{PN}_i^{\mathchorus{C}}})$) to allow faster execution. The normal tasks based requests are sorted as per the basis of their priorities and the physical nodes are assorted  in decreasing order of their processing speed in the list. Such an allocation of virtual nodes executes lowest deadline requests on highest processing speed servers to minimize the time of execution. The constraint $C_1$ specifies $k^{th}$ task can be assigned to only one $\mathchorus{VN}_j$ hosted on one $\mathchorus{PN}_i$ at an instance;  \{$C_2$-$C_4$\} state resource capacity requirement (viz., CPU ($\mathchorus{C}$), memory ($\mathchorus{M}$), bandwidth ($\mathchorus{B}$)) of $VN_j$ must be lesser or equal to available resource capacity ($\mathchorus{C}^{\ast}$), memory ($\mathchorus{M}^{\ast}$), bandwidth ($\mathchorus{B}^{\ast}$)) of $\mathchorus{PN}_i$; and $C_5$ \& $C_6$ specify  resource requirement ($R$) of $k^{th}$ job ($\lambda_k$) for processing must be satisfied by the available resources ($R^{\ast}$) on the respective $\mathchorus{VN}_j$ hosted on $\mathchorus{PN}_i$. 
\begin{gather}
\omega_{kji}^{\mathds{C}^{St}_k}= \mathchorus{VN}_j^{R^{\mathds{C}^{St}_k}} \times MAX(List_{\mathchorus{PN}_i^{\mathchorus{M}}}) \quad s.t.\{C_1-C_6\} \label{re1}\\	
\omega_{kji}^{\mathds{C}^{Rh}_k}= \mathchorus{VN}_j^{R^{\mathds{C}^{Rh}_k}} \times MAX(List_{\mathchorus{PN}_i^{R} })\quad s.t.\{C_1-C_6\} \label{re2}\\
\omega_{kji}^{\mathds{C}^{Nt}_k}= \mathchorus{VN}_j^{R^{\mathds{C}^{Nt}_k}} \times MAX(List_{\mathchorus{PN}_i^{\mathchorus{C}}}) \quad s.t.\{C_1-C_6\} \label{re3}
\end{gather}
\begin{equation}\label{cs}
\centering
\begin{aligned}
\left.
\begin{array}{ll}
{C_1:} \quad {\omega_{kji}}\mapsto \mathchorus{PN}_i \times \mathchorus{VN}_j \times \lambda_k = 1  \quad \forall_{k \in m}\forall_{j \in Q}\forall_{i \in P}\\
C_2: \quad {\mathchorus{VN}_j^\mathchorus{C}} \times \omega_{kji} \leq \mathchorus{PN}_i^{\mathchorus{C}^{\ast}}  \quad \forall_{k \in m}\forall_{j \in Q}\forall_{i \in P}\\
C_3: \quad {\mathchorus{VN}_j^{\mathchorus{M}}} \times \omega_{kji} \leq \mathchorus{PN}_i^{\mathchorus{M}^{\ast}}  \quad \forall_{k \in m}\forall_{j \in Q}\forall_{i \in P}\\
C_4: \quad {\mathchorus{VN}_j^{\mathchorus{B}}} \times \omega_{kji} \leq \mathchorus{PN}_i^{\mathchorus{B}^{\ast}}\\
C_5: \quad  \sum_{k\in M} \lambda_k \times {R}_k  < \sum_{i \in P}\mathchorus{PN}_i^{{R^{\ast}}} \quad R^{\ast} \in \{\mathchorus{C}^{\ast},\mathchorus{M}^{\ast}, \mathchorus{B}^{\ast}\} \\
C_6: \quad  \lambda_k \times {R}_k \leq \mathchorus{VN}_j^{R^\ast} \quad \forall_k \in [1, m], j \in [1, Q] \\
\end{array}
\right\}  
\end{aligned}
\end{equation}

\subsection{Reliability Modeling} \label{rem}
Let $\mathchorus{PN}_i^{\mathchorus{C}}$, $\mathchorus{PN}_i^\mathchorus{M}$ and $\mathchorus{PN}_i^\mathchorus{B}$ are CPU, memory, and bandwidth capacity of $i^{th}$ physical node. Simultaneously, $\mathchorus{VN}_j^\mathchorus{C}$, $\mathchorus{VN}_j^\mathchorus{M}$ and $\mathchorus{VN}_j^\mathchorus{B}$ depict CPU, memory, and bandwidth utilization, respectively for $j^{th}$ virtual node. A mapping $\omega$ represents allocation of $Q$ virtual nodes among $P$ physical nodes, wherein $\omega_{ji}=1$, if server $\mathchorus{PN}_i$ hosts $\mathchorus{VN}_j$, else it is $0$. The usage of each resource is separately monitored where, CPU, memory and bandwidth utilization of a server can be calculated using Eqs. (\ref{cpu})-(\ref{ram}). Accordingly, Eq. (\ref{cpu1}) evaluates resources utilization of $i^{th}$ server i.e., ($\mathchorus{RU}_{i}^{{R}}$) is determined by applying Eq. (\ref{cpu1}); where $Num$ is the number of resources. The essential reliability attributes considered in the proposed approach includes \textit{availability}, \textit{fault tolerance}, and \textit{failure detection}. The global reliability ($\mathds{R}_i$) of $i^{th}$ physical node is computed using Eq. (\ref{rm1}), where $\mathchorus{R}a$ is a reliability level associated with a specific attribute. The failure rate or hazard ($\mathchorus{H}z_{ji}$) of a virtual node ($\mathchorus{VN}_j$) deployed on $i^{th}$ physical node ($\mathchorus{PN}_i$) having resource utilization $\mathchorus{RU}_i$ is given by Eq. (\ref{rm2}),  where $ {\mathchorus{H}z}_{i}^{MAX}$ is the hazard rate of physical node ($\mathchorus{PN}_i$) at maximum resource utilization.

\begin{gather}
\mathchorus{RU}_i^\mathchorus{C}=\frac{\sum_{j=1}^{Q}{\omega_{ji} \times \mathchorus{VN}_j^\mathchorus{C}}}{\mathchorus{PN}_i^{C}}\label{cpu}\\
{\mathchorus{RU}_i}^\mathchorus{M}=\frac{\sum_{j=1}^{Q}{\omega_{ji} \times \mathchorus{VN}_j^\mathchorus{M}}}{\mathchorus{PN}_i^\mathchorus{M}}\label{memory}\\
{\mathchorus{RU}_i}^\mathchorus{B}=\frac{\sum_{j=1}^{Q}{\omega_{ji} \times \mathchorus{VN}_j^\mathchorus{B}}}{\mathchorus{PN}_i^\mathchorus{B}}\label{ram}\\
\mathchorus{RU}_{i}= \frac{\mathchorus{RU}_i^\mathchorus{C} + \mathchorus{RU}_i^\mathchorus{M} +\mathchorus{RU}_i^\mathchorus{B}}{|{Num}|} \label{cpu1} 
\end{gather}

\begin{gather}
\mathds{R}_i= \Pi_{a=1}^{4}{\mathchorus{R}a} \label{rm1}\\
{\mathchorus{H}z}_{ji}= {\mathchorus{H}z}_{i}^{MAX} \times \mathchorus{RU}_i^{\gamma} \label{rm2}
\end{gather}
The term $\gamma$ defines the `sensitivity factor' which determines the sensitivity of the failure rate towards the resource utilization and $\gamma=1$ represents a linear relationship between $\mathchorus{H}z_{ji}$ and $\mathchorus{RU}_i$. The maximum hazard rate (${\mathchorus{H}z}_{j}^{MAX}$) of physical node ($\mathchorus{PN}_i$) in terms of mean time between failure (MTBF) is given by Eq. (\ref{rm3}). All the virtual nodes  hosted on a physical node ($\mathchorus{PN}_i$) fail with the failure of $\mathchorus{PN}_i$. Hence, there is a linear relationship among the physical and virtual nodes. The reliability of  physical node ($\mathchorus{PN}_i$) while running $q^\ast$ virtual nodes is computed using Eq. (\ref{rm4}). The total reliability of the system ($\mathds{R}_{Sys}$)  over time-interval $t+ \Delta t$ is computed using Eq. (\ref{rm5}).
\begin{gather}
{\mathchorus{H}z}_{i}^{MAX}= \frac{1}{MTBF} \label{rm3}\\
\mathds{R}_i  = \int_{t}^{t + \Delta t}\Pi_{j=1}^{q^\ast}(exp^{-\mathchorus{H}z}_{jk} \times L_{Max}) dt \label{rm4}\\
\mathds{R}_{Sys}  = \int_{t}^{t + \Delta t}\Pi_{i=1}^{P} ({\mathds{R}_i })dt \label{rm5}
\end{gather}

\subsection{Energy Efficiency Modeling} \label{eem}

In the proposed model, the power consumption of all the physical nodes is based on inbuilt Dynamic Voltage Frequency Scaling (DVFS) energy saving technique \cite{minas2009energy} that considers sleep and busy states  of CPU. During sleep state, CPU works with reduced clock cycle and some internal components of CPU are in non-operational mode while in the busy state, power consumption depends on the CPU utilization rate. Hence, the power consumption for $i^{th}$  physical node and entire system $Sys$ for time-interval \{$t$, $t + \Delta t$\} is described by the linear functions stated in Eq. (\ref{power1}) and (\ref{power2}), respectively, where $\mathchorus{RU}_{i} \in$ {[0, 1]} is resource utilization of $i^{th}$ physical node,  ${\varepsilon(\mathchorus{RU})_i}^{Max}$, ${\varepsilon(\mathchorus{RU})_i}^{Min}$, and ${\varepsilon(\mathchorus{RU})_i}^{Idle}$ are maximum, minimum, and idle state power consumption of physical nodes $\mathchorus{PN}_i$.  The total energy consumption ($E$) of a server is computed using Eq. (\ref{p1});  where $\mathchorus{RU}$ is the resource utilization of the server and $\varepsilon(\mathchorus{RU})$ is the rate of energy consumption.  For time-interval	 \{$t$, $t + \Delta t$\}, the expected energy consumption of the entire system ($Sys$)  is obtained by applying Eq. (\ref{energy}).
\begin{gather}
\varepsilon(\mathchorus{RU})_i =([{ \varepsilon(\mathchorus{RU})_i}^{Max} - { \varepsilon(\mathchorus{RU})_i}^{Min}] \times \mathchorus{RU}_{i} + { \varepsilon(\mathchorus{RU})_i}^{Idle}) \label{power1}\\
\varepsilon(\mathchorus{RU})_{{Sys}} = \int_{t}^{t + \Delta t}
\bigg(\sum_{i=1}^{P} { \varepsilon(\mathchorus{RU})_i}\bigg){dt}
\label{power2}
\end{gather} 

\begin{equation} \label{p1}
E=\int_{t}^{t + \Delta t}	{ \varepsilon(\mathchorus{RU}){dt}}
\end{equation}

\begin{equation} \label{energy}
E_{Sys}= \varepsilon(\mathchorus{RU})_{Sys} \times \Delta t
\end{equation}

\section{Workload Estimation} \label{qmlwa}
This section entails  the architecture of the proposed QML driven artificial neural network model and its operational process for a proactive estimation of the heterogeneous workloads (Fig. \ref{fig:proposed-model}, BLOCK 2). A \textit{Toffoli Gate embedded Quantum Neural Network} (TG-QNN) is developed, comprising of $n$,  $p$, and  $q$  qubit neurons \{$\Phi_1$, $\Phi_2$, ..., $\Phi_S$\} in  input, multiple hidden, and output layers, respectively. These layers are inter-connected via qubit weights \{$\Theta^I_1$, $\Theta^I_2$, ..., $\Theta^H_1$, $\Theta^H_2$, ..., $\Theta^O_1$, ..., $\Theta^O_q$\} with  size of the neural network $S$  such that $S=(n \times p) + d \times (p \times  p) + (p \times q)$), where $d$ is the number of hidden layers. The live resource usage samples \{$\mathchorus{D}_1$, $\mathchorus{D}_2$, ..., $\mathchorus{D}_N$\} $\in \mathchorus{D}$ are transformed into  qubit input vector \{$\Phi^I_1$, $\Phi^I_2$, ..., $\Phi^I_N$\}$\in \mathchorus{X}^{\ast}$ with the help of Quantum Rotation gate  using Eq. (\ref{qubit}) to be fed as a training data samples into the input layer of TG-QNN. The $N$ previous workload values are analysed to extract
critical behavioral patterns from actual workload samples to forecast the approaching workload (i.e., ${N+1}^{th}$) information.
\begin{gather} \label{qubit}
\Phi_i = \frac{\pi}{2} \times \mathchorus{D}_i	
\end{gather}
The learning process of TG-QNN is achieved by implementing and utilizing  \textit{Quantum version of BlackHole Optimization} (i.e., QBHO) algorithm which selects the most suitable qubit network from the random population of $\mathchorus{Z}$ qubit networks \{$\Theta^\ast_1$, $\Theta^\ast_2$, ..., $\Theta^\ast_\mathchorus{Z}$\}.  The algorithm iteratively optimizes the qubit neural weights by exploring and exploiting  the diverse population of qubit networks and selects the best qubit network candidate by evaluating a fitness function ($Eval$()) computing prediction error. TG-QNN incorporates non-linearity element at each neuron of all the layers (except input layer) by executing three consecutive steps: \textit{Qubit summation}, \textit{Controlled reverse operation}, and \textit{Activation}. These steps activates the qubit network learning process with greater flexibility by producing relevant, complex, and intuitive patterns. The Hidden layer neurons produce qubit vector \{$\Phi^{H_i}_1$, $\Phi^{H_i}_2$, ..., $\Phi^{H_i}_p$\} such that $1 \leq i \leq d$ by computing Eqs. (\ref{h1})-(\ref{h3}):
\begin{gather}  
\varpi_j^{H_i}=\sum_{i=1,j=1}^{n, {p}} \mathchorus{F}(\Theta_{ij}^{I})\times \mathchorus{F}( \Phi_i^{I})-\mathchorus{F}(\Omega^{\mathchorus{B}})  \label{h1} \\
\Phi_j^{H_i} = \frac{\pi}{2}\times \mathchorus{F}(Z_1 \oplus Z_2) - arg(\varpi_j^{H_i}) \label{h2}\\   
\mathchorus{Y}_j^{H_i} =\mathchorus{F}(\Phi_j^{H_i}) \label{h3}
\end{gather}
where, $\varpi_j^{H_i}$ represents qubit summation (Eq. (\ref{h1})) obtained for the  $j^{th}$ neuron of $i^{th}$ hidden layer. The adjusted qubit vector ($\Phi_j^{H_i}$) is obtained by applying controlled-controlled NOT effect of the Toffoli Gate using Eq. (\ref{h2}), where $Z_1$ and $Z_2$ are controlled parameters such that the value for $Z_1$ is randomly generated either 0 or 1 while the value of $Z_2$ is governed by the comparative fitness value of current and previous candidate solutions.  
The expression: $\mathchorus{F}(\Phi_j^{H_i})$ represents the  $j^{th}$ adjusted qubit vector obtained by applying activation function to $\Phi_j^{H_i}$ and Eq. (\ref{h3}) generates the final outcome $\mathchorus{Y}_j^{H_i}$  of $j^{th}$ neuron of the $i^{th}$ hidden layer. On the same lines, an output qubit vector \{$\Phi^{O}$\} is generated at output layer by computing Eqs. (\ref{o1})-(\ref{o3}), wherein $\varpi_j^{O_i}$ is  the qubit summation (Eq. (\ref{o1})); the effect of Toffoli Gate on $\varpi_j^{O_i}$ is generated as  $\Phi_j^{O}$  by applying Eq. (\ref{o2}). The final outcome of  $j^{th}$ neuron ($\mathchorus{Y}_j^{O_i}$) is obtained by executing the activation operation using Eq. (\ref{o3}).
\begin{gather}  
\varpi_j^{O_i}=\sum_{i=1,j=1}^{Q^{p}, Q^q} \mathchorus{F}(\Theta_{ij}^{H})\times \mathchorus{F}( \Phi_i^{H})  \label{o1} \\
\Phi_j^{O_i} = \frac{\pi}{2}\times \mathchorus{F}(Z_1 \oplus Z_2) - arg(\varpi_j^{O_i}) \label{o2}\\   
\mathchorus{Y}_j^{O_i} =\mathchorus{F}(\Phi_j^{O_i}) \label{o3}
\end{gather}
The activation function $\mathchorus{SIG}()$ given in Eq. (\ref{sigmoidfunction}) admits non-linearity and maps input to output in the range [0, 1]. Therefore, the final predicted  values are obtained in this normalized range to make them comparable with the actual output values. The performance and accuracy of the prediction model are evaluated using the mean squared error score  ($\mathchorus{MSE}$) using Eq. (\ref{rmse}) for $m$  data samples. 
\begin{gather}
\mathchorus{SIG}(\Phi)=\frac{1}{1+exp^{(-\Phi)}}\label{sigmoidfunction}\\
\mathchorus{MSE} = {\frac{1}{m}\sum_{i=1}^{m}(\mathchorus{D}_i^{Ac}-\mathchorus{D}_i^{Pr})^2}\label{rmse} 
\end{gather}

\subsection{TG-QNN Optimization by QBHO} \label{Qdpbho}
The  learning process initializes with generation of random population of candidate solutions or \textit{qubit stars} using Eqs. (\ref{eqn:qq}) and (\ref{eqn:qq1}), where $\alpha=\sqrt{(rd)}$ and $\beta=\sqrt{(1-rd)}$, $rd$ is a random number in the  range [0, 1], $\alpha$ and $\beta$ are the probability amplitudes for realizing $\ket{0}$ and $\ket{1}$ respectively, $\Psi$ is a quantum state in the Hilbert vector space and $\Theta$ is the corresponding qubit state. The qubit star with the best fitness value is choosen as a `\textit{Quantum Black-hole}'.  As depicted in Fig. \ref{fig:proposed-model} BLOCK 2, the QBHO algorithm involves three consecutive stages:  \textit{Qubit cluster optimization}, \textit{Heuristic optimization}, and  \textit{Position Update}.

\subsubsection{Qubit cluster optimization}
The $\mathchorus{Z}$ qubit stars are organized into $\mathchorus{K}$  clusters of qubits by applying Eq. (\ref{eq.cluster}), where $\mathchorus{W}_{ik}$ defines mapping of $i^{th}$ Qubit network ($\Theta^\ast_i$) and $\mu_k$ is  the centroid of $k^{th}$ Qubit Cluster ($QC$).  The inter-clusters are distinguished on the basis of similarity of amplitude of the qubit weights within a quantum network such that the qubit clusters are  distinguishable. All the candidates of each qubit cluster  are examined  using  Eq. (\ref{rmse}) over training data. The best solution of each $k^{th}$ cluster  is selected as a local blackhole ($QC_{k}^{best}$), wherein $QC_{k}^{best} =$ BEST(\{$\Theta^\ast_1$, $\Theta^\ast_2$, ..., $\Theta^\ast_{\mathchorus{Z}/\mathchorus{K}}$\}). 
\begin{gather}
\label{eqn:qq}
\vartheta=\alpha + i\beta \\
\label{eqn:qq1}
\Theta= phase(\vartheta)\times \frac{\pi}{2}  
\\ 
\label{eq.cluster}
\mathchorus{G}=\sum_{i=1}^{\mathchorus{Z}}\sum_{k=1}^{\mathchorus{K}}{\mathchorus{W}_{ik}{|{\Theta^\ast}_i- \mu_k|}^2}
\end{gather}

\subsubsection{Heuristic optimization}
The qubit blackholes ($QC_i^{best}$: $0 \leq i \leq \mathchorus{K}$) consitute the  population for next stage, where \textit{heuristic crossover} operation  produces new candidates of superior efficiency. The two qubit stars selected randomly act as parent chromosomes and their fitness is evaluated using Eq. (\ref{rmse}) and compared to locate  the best parent chromosome.  Thereafter, the updated candidate solutions are produced using Eq. (\ref{heuristic})  by combining the two $QC^{best}$ blackholes \{$QC_{k}^{best}$ and $QC_{j}^{best}$\},  where $k \neq j$, $i \in [1, \mathchorus{Z}/\mathchorus{K}]$, $\mathchorus{CR}_i$ is a crossover rate randomly generated in the range [0, 1] for $i^{th}$ gene. Likewise, $\mathchorus{K}$ new offsprings are produced i.e., one from each member of the population of qubit cluster blackholes ($QC^{best}$). The best candidate nominated as `\textit{Global Best}' ($QG^{best}$) is selected using Eq. (\ref{g2}) to enhance the diversity of the qubit cluster population by inducing qubit network candidates with improved fitness values.
\begin{gather}
\label{heuristic}
QG_k^{Off} = \mathchorus{CR}_i \times (QC_{k}^{best}-QC_{j}^{best}) + QC_{k}^{best} 
\\
\label{g2}
QC^{best}_{j}=\begin{cases}
QG_k^{Off} & \text{If ($\mathchorus{F}$($QG_k$) $\geq$ $\mathchorus{F}$($QC^{best}_{j}$))} \\
QC^{best}_{j} & {\text{Otherwise}} 
\end{cases}
\end{gather}

\subsubsection{Position Update}
The position of qubit stars update according to the values of $QC_{k}^{best}$ and $QG^{best}$ using Eq. (\ref{bl1}), where ${\Theta^\ast}_i^k(t)$ and ${\Theta^\ast}_i^k(t+1)$ 
are the positions of $i^{th}$ star of $k^{th}$ sub-population at instances $t$ and $t+1$, respectively. The terms $r_1$ and $r_2$ are random
numbers in the range [0, 1]  while $\gamma_c$
and $\gamma_h$ are the attraction forces applied on ${\Theta^\ast}_i^k(t)$
by $QC_{k}^{best}$ and $QG^{best}$, respectively. Eq. (\ref{bl1}) employs
$QC_{k}^{best}$ and $QG^{best}$ for changing the positions of the qubit stars to  maintain the diversity of the population  by  controlling the convergence speed and maintaining the optimum level of exploration.
\begin{equation}\label{bl1}
\begin{aligned}
\mathchorus{CF}(t)= \gamma_c^k r_1\big({QC}_{k}^{best}(t)-{\Theta^\ast}^i_k(t) \big)\\
\mathchorus{HF}(t)= \gamma_h r_2\big({QG}^{best}(t) - {\Theta^\ast}^i_k(t)\big) \\
{\Theta^\ast}_i^k(t+1)= {\Theta^\ast}_i^k(t)+\mathchorus{CF}(t) +\mathchorus{HF}(t)
\end{aligned}
\end{equation}

All the updated qubit stars are evaluated using Eq. (\ref{rmse}) to locate a better solution (qubit network) within $k^{th}$ qubit cluster. Accordingly, the values of $QC_k^{best}$ and $QG^{best}$ are relocated.  QBHO algorithm prohibits the entry of the candidate solution returning back from an event horizon ($\mathchorus{EH}$) area of a blackhole solution which is marked-out by the radius of event horizon ($\mathchorus{R}_\mathchorus{EH}$). The ratio of the fitness value of a qubit cluster blackhole ($\mathchorus{F}(QC^{best})$) and  summation of the fitness values of its sub-population ($\sum_{i=1}^{\mathchorus{Z}/\mathchorus{K}}{\mathchorus{F}({{\Theta^\ast}^i_k})}$) computes an event horizon radius ($\mathchorus{R}_{\mathchorus{EH}}$) for the respective blackhole using Eq. (\ref{r1}). Similarly, the event horizon radius of a global blackhole ($\mathchorus{R}_{{EH}}\big(QG_{k}^{best}\big)$)  is evaluated using Eq. (\ref{r2}), where $\mathchorus{F}(QG_{k}^{best})$ is fitness value of  heuristic global blackhole, $\sum_{k=1}^{K}\sum_{i=1}^{\mathchorus{Z}/\mathchorus{K}}{\mathchorus{F}({\Theta^\ast}_{k}^{i})}$ is a fitness value of the entire population. 
\begin{gather} \label{r1}
\mathchorus{R}_{\mathchorus{EH}}\big(QC_{k}^{best}\big)=\frac{\mathchorus{F}(QC_{k}^{best})}{\sum_{i=1}^{\mathchorus{Z}/\mathchorus{K}}{\mathchorus{F}({\Theta^\ast}_{k}^{i})}} \quad k \in [1, K]
\\ \label{r2}
\mathchorus{R}_{\mathchorus{EH}}\big(QG^{best}\big)=\frac{\mathchorus{F}(QG_{k}^{best})}{\sum_{k=1}^{\mathchorus{K}}\sum_{i=1}^{\mathchorus{Z}/ \mathchorus{K}}{\mathchorus{F}({\Theta^\ast}_{k}^{i})}} 
\end{gather}
The distance between the best and rest of the solutions is determined by computing
the arithmetic difference of their fitness values to confirm that a member solution
has entered into an event horizon of the blackhole solution. Each qubit star attracts toward the qubit cluster and  heuristic global blackholes and their corresponding distances from  these two blackholes are computed  using  Eqs. (\ref{d1}) and  (\ref{d2}):
\begin{gather}
\resizebox{0.43\textwidth}{!}{$ 
	\mathchorus{D}_{QC_{k}^{best}}\big(\Theta_{k}^{i}\big)=\mathchorus{F}(QC_{k}^{best}) - \mathchorus{F}(\Theta_{k}^{i}) \quad i \in [1, \mathchorus{Z/K}]  $}\label{d1}
\\ \label{d2}
\mathchorus{D}_{QG_{k}^{best}}\big(\Theta_{k}^{i}\big)=\mathchorus{F}(QG_{k}^{best}) - \mathchorus{F}(\Theta_{k}^{i}) \quad i \in [1, 2\mathchorus{K}]
\end{gather}
wherein $\mathchorus{D}_{QC_{k}^{best}}\big(\Theta_{k}^{i}\big)$ and $\mathchorus{D}_{QG_{k}^{best}}\big(\Theta_{k}^{i}\big)$ are the distances of $i^{th}$ qubit star ($\Theta_{k}^{i}$) of $k^{th}$ cluster from qubit cluster blackhole ($QC_{k}^{best}$) and global blackhole ($QG^{best}$), respectively.

{In QBHO, the event horizon is mathematically modeled as the ratio of a blackhole’s fitness to the total fitness of its population, determining a dynamic radius around the best solution. Qubit stars within this radius are absorbed and reinitialized, enhancing convergence speed while preserving diversity and preventing premature stagnation. This mechanism ensures that the population is steadily guided toward optimal regions while maintaining exploration, leading to faster and more reliable convergence.}
The operational summary of QBHO is depicted in Algorithm \ref{algo-dpbho}. 
\begin{figure}[!htbp]
	\removelatexerror
	\begin{algorithm}[H]
		\caption{TG-QNN Learning by QBHO}
		\label{algo-dpbho}
		{Initialize $N$ random solutions: \{${\Theta}_1$, ${\Theta}_2$, ..., ${\Theta}_{Z}$\}$\in {X}$ \;	
			Organize ${X}$ into ${K}$ Qubit clusters and evaluate each random solution on training data using Eq. (\ref{rmse}) \;
			\For{each k = \{1, 2, ..., ${K}$\}}{$QC_{k}^{best}$ = Best(\{${\Theta}_1$, ${\Theta}_2$, ..., ${\Theta}_{{Z}/ {K}}$\})\; }
			\For{ i= \{1, 2, ..., ${K}$\}}{Two parents ($QC_{k}^{best}$ and $QC_{j}^{best}$) are randomly selected and a new offspring ($QG_{k}^{Off}$) is produced using Eq. (\ref{heuristic})\;
				Evaluate $QG^{Off}_{k}$ over training data using Eq. (\ref{rmse}) \;
				\If{fitness value of $QG^{Off}_{k} \geq$  less best parent ($QC_{j}^{best}$) }{
					Replace  $QC_{j}^{best}$ with $QG^{Off}_{k}$ \; 
				}
			}
			\For{each k = \{1, 2, ..., ${K}$\}}{$QG^{best}$ = Best(\{$QC_1^{best}$, $QC_2^{best}$, ..., $QC_{{K}}^{best}$\}) \; }
			\While{termination }{
				Update position of ${\Theta}^i_k$ using Eq. (\ref{bl1})\;
				Evaluate ${\Theta}^i_k (t+1)$ using Eq. (\ref{rmse}) \;
				\For{each k = \{1, 2, ..., ${K}$\}}{$QC_{k}^{best}(t+1)$ = Best($QC_{k}^{best}(t)$, \{${\Theta}_1$, ${\Theta}_2$, ..., ${\Theta}_{{Z}/ {K}}$\})\; 
				}
				$QG_{best}(t+1)$ = Best($QG_{best}(t)$, \{$QC_1^{best}$, $QC_2^{best}$, ..., $QC_{{K}}^{best}$\}) \;	
				Estimate the radius and distances using Eqs. (\ref{r1}-\ref{d2}) \;
				\For{each i = \{1, 2, ..., ${Z}$\}}{
					\If{(${D}is_{QC_{k}^{best}}\big(\Theta_{k}^{i}\big) \leq $
						${R}_{{EH}}\big(QC_{k}^{best}\big)$) $\vee$ (${D}is_{QG^{best}}\big(\Theta_{k}^{i}\big) \leq $
						${R}_{{EH}}\big(QG^{best}\big)$) }{Collapse ${\Theta}^i_k$ and rejuvenate population by adding a new candidate using Eqs. (\ref{eqn:qq}) and (\ref{eqn:qq1})\;}

				}		
			}		
	} 	\end{algorithm}
	
\end{figure}
\\
\textit{Time Complexity}: Step 1 initializes random solutions, has complexity $\mathcal{O}(1)$. Step 2 trains TG-QNN and evaluates ${M}$ samples for fitness of ${Z}$ solutions across network: $({n} \times p) + (p \times q)$, wherein $n > p$ and  $q=1$ producing complexity: $\mathcal{O}({Z} \times {n}^2 \times {M})$. Steps [3-5], steps [6-12], and steps [13-15] iterate ${K}$ times have equal time complexity of $\mathcal{O}({K})$. Assume steps [16-29] repeat for $t$ intervals, wherein steps [19-21] have $\mathcal{O}(\mathcal{K})$  while steps [24-28] have $\mathcal{O}({Z})$ complexities. Hence, the total time complexity for QBHO algorithm is $\mathcal{O}({Z}\times n^2 \times {M} \times {K})$.

\section{Traffic States  Entropy Analysis} \label{tscea}
TSECE computes Eqs. (\ref{e1}-\ref{delta}) to analyse $n$ states of traffic \{$\mathds{TS}_1$, $\mathds{TS}_2$, ..., $\mathds{TS}_n$\} for a specific duration ($\mathchorus{D}$) by comparing the estimated ($\mathchorus{TW}^{Pr}_{t-1}$) and actual ($\mathchorus{TW}^{Ac}_t$) resource  utilization of the previously predicted workload and actual traffic, respectively,  at $t^{th}$ time instance. The trail ($\Psi$) between actual and predicted traffic workloads  due to resources \{CPU ($\mathchorus{C}$),  memory ($\mathchorus{M}$), and bandwidth ($\mathchorus{B}$)\} utilization associated with physical nodes is estimated using Eqs. (\ref{e1}-\ref{e3}). The term $\uplus$ evaluates the lag between threshold duration ($\mathchorus{D}^{thr}$) and execution time ($\mathchorus{ED}$) of $\mathchorus{TW}^{Pr}_t$ and $\mathchorus{TW}^{Ac}_t$,  respectively. The value of threshold duration is not fixed and may change as per the demand of the workload execution.  However, the value of $\uplus$ helps in deciding the traffic condition such that $\uplus_t < 0$ detects probable `network congestion' while $\uplus_t \ge 0$ specifies `regular traffic'. 
\begin{gather} \label{e1}
\Psi^{\mathchorus{C}}_t=\mathchorus{TW}^{Pr}_{t}\times \mathchorus{C}^{Pr} -\mathchorus{TW}^{Ac}_{t}\times \mathchorus{C}^{Ac}
\\ \label{e2}
\Psi^{\mathchorus{M}}_t=\mathchorus{TW}^{Pr}_{t}\times \mathchorus{M}^{Pr} -\mathchorus{TW}^{Ac}_{t}\times \mathchorus{M}^{Ac}
\\ \label{e3}
\Psi^{\mathchorus{B}}_t=\mathchorus{TW}^{Pr}_{t}\times \mathchorus{B}^{Pr} -\mathchorus{TW}^{Ac}_{t}\times \mathchorus{B}^{Ac}
\\ \label{delta}
\uplus_t =(\mathchorus{TW}^{Pr}_{t} \times \mathchorus{D}^{thr})-(\mathchorus{TW}^{Ac}_{t} \times \mathchorus{ED})
\end{gather}

Eq. (\ref{states}) determines various states of live internet traffic: \{$\mathds{TS}_1$, $\mathds{TS}_2$, ..., $\mathds{TS}_{10}$\} at $t^{th}$ instance by considering the values of $\Psi^x_t$ ($x \in \{\mathchorus{C}, \mathchorus{M}, \mathchorus{B}\}$) and  $\uplus_t$ obtained via Eqs. (\ref{e1}-\ref{e3}) and  Eq. (\ref{delta}), respectively.  The magnitude of different states $n$ (n=10 for the experimental evaluation of the  proposed work) states viz., $\mathds{TS}_1$, $\mathds{TS}_2$, ..., $\mathds{TS}_9$, $\mathds{TS}_{10}$ in TSECE unit is computed in terms of $\Omega_z: z=\{1, n\}$ using Eq. (\ref{omega}). The mapping $\sum_{i=1
}^{Q} |\mathchorus{VN}_i \times \mathchorus{TW}^{Ac}_{i} \times \uplus_i  \times \Psi^R_i|$ measures the quantity of virtual nodes associated with $z^{th}$ traffic states on the basis of the values of $\mathchorus{TW}^{Ac}_{i}$, $\uplus_i$, $\Psi_i$ for virtual nodes $\mathchorus{VN}_i$, and total number of virtual nodes ($|\mathds{VN}|$).        

\begin{equation}\label{states}
\resizebox{0.48\textwidth}{!}{$  
\mathds{TS}_z=	\begin{cases}
\mathds{TS}_1, & {If(\mathchorus{TW}^{Ac}_{t}\times \uplus_t \times \Psi^\mathchorus{C}_t \quad \&\& \quad \Psi^\mathchorus{C}_t>0, \uplus_t>0)} \\
\mathds{TS}_2, & {If(\mathchorus{TW}^{Ac}_{t}\times \uplus_t \times \Psi^\mathchorus{C}_t \quad \&\& \quad \Psi^\mathchorus{C}_t<0, \uplus_t>0)} \\
\mathds{TS}_3, & {If(\mathchorus{TW}^{Ac}_{t}\times \uplus_t \times \Psi^\mathchorus{C}_t \quad \&\& \quad \Psi^\mathchorus{C}_t \sim 0, \uplus_t>0)} \\
\mathds{TS}_4, & {If(\mathchorus{TW}^{Ac}_{t}\times \uplus_t \times \Psi^\mathchorus{M}_t \quad \&\& \quad \Psi^\mathchorus{M}_t>0, \uplus_t>0)} \\
\mathds{TS}_5, & {If(\mathchorus{TW}^{Ac}_{t}\times \uplus_t \times \Psi^\mathchorus{M}_t \quad \&\& \quad \Psi^\mathchorus{M}_t<0, \uplus_t>0)} \\
\mathds{TS}_6, & {If(\mathchorus{TW}^{Ac}_{t}\times \uplus_t \times \Psi^\mathchorus{M}_t \quad \&\& \quad \Psi^\mathchorus{M}_t \sim 0, \uplus_t>0)} \\
\mathds{TS}_7, & {If(\mathchorus{TW}^{Ac}_{t}\times \uplus_t \times \Psi^\mathchorus{B}_t \quad \&\& \quad \Psi^\mathchorus{B}_t>0, \uplus_t>0)} \\
\mathds{TS}_8, & {If(\mathchorus{TW}^{Ac}_{t}\times \uplus_t \times \Psi^\mathchorus{B}_t \quad \&\& \quad \Psi^\mathchorus{B}_t<0, \uplus_t>0)} \\
\mathds{TS}_9, & {If(\mathchorus{TW}^{Ac}_{t}\times \uplus_t \times {\Psi^\mathchorus{B}}_t \quad \&\& \quad \Psi^\mathchorus{B}_t \sim 0, \uplus_t>0)} \\
\mathds{TS}_{10} & {\text{otherwise}}   
\end{cases}$}
\end{equation}

\begin{equation} \label{omega}
\Omega_x=\frac{\sum_{i=1
	}^{Q} |\mathchorus{VN}_i \times \mathchorus{TW}^{Ac}_{i} \times \uplus_i  \times \Psi^R_i|}{|\mathds{VN}|}\quad \forall_{\Psi_i>0, \uplus_i>0}
\end{equation}

The  entropy of the live internet traffic ($\mathds{E}n_t$) is computed using Eq. (\ref{entropy1}). The entropy of the forthcoming traffic (i.e., $\mathds{E}n_{t+1}$) is quantified by evaluating Eq. (\ref{entropy2}) which  analyses   previous stages entropy such as: $\mathds{E}n_{t-1}$ and $\mathds{E}n_{t-2}$ as well as current ($\xi_t$) traffic entropy. The terms $\xi$ and $\xi^{\ast}$ assign weight to the entropy in the range [0, 0.2], where more weight is allotted to the most recent traffic entropy such that $\mathds{E}n_{t} >  \mathds{E}n_{t-1} \geq \mathds{E}n_{t-2} $ to estimate the entropy  of the approaching traffic  for an efficient workload distribution.
\begin{equation} \label{entropy1}
\mathds{E}n_t=\sum_{i=1}^{n} {-\mathds{TS}_i log_2(\mathds{TS}_i)}
\end{equation}
\begin{equation}\label{entropy2}
\mathds{E}n_{t+1}=\xi^{\ast} \times \mathds{E}n_{t-2} + \xi \times \mathds{E}n_{t-1} + (1-\xi-\xi^{\ast})\times \mathds{E}n_t
\end{equation}

The number of required physical nodes is enumerated using Eq. (\ref{lm}) which incorporates           $\mathchorus{TW}^{Pr}_{t+1}$ and $\mathds{E}n_{t+1}$ such that combined resource utilization ($\mathchorus{RU} $) of $\mathchorus{TW}^{Pr}_{t+1}$ and $\mathds{E}n_{t+1}$ must be lesser than the total resource capacity ($\mathchorus{R}$) of the physical nodes within the cluster. Eqs. (\ref{ac}) and (\ref{inac}) calculate the number of active physical nodes ($\mathchorus{PN}_{\mathchorus{AC}}$) and inactive ($\mathchorus{PN}^{}_{\mathchorus{IN}}$) physical nodes within the cluster of $P$ physical nodes, where \{$P^\ast, P^{\ast\ast}$\}$\in P$.  
\begin{gather}\label{lm}
\mathchorus{RU} \times (\mathchorus{TW}^{Pr}_{t+1} + \mathds{E}n_{t+1}) \le \sum_{i=1}^{K}{\mathchorus{PN}_i \times {\mathchorus{R}}}	
\\
\label{ac}
\mathchorus{PN}^{}_{\mathchorus{AC}}: \sum_{i=1}^{P^\ast}|{\mathchorus{PN}^{\mathchorus{R}}_i} \ge \mathchorus{TW}^{Pr}_{t+1}\times \mathchorus{RU}^{Pr}|   	
\\
\label{inac}
\mathchorus{PN}^{}_{\mathchorus{IN}}: \sum_{i=1}^{P^{\ast\ast}}|{\mathchorus{PN}^{\mathchorus{R}}_i} \ge \mathds{E}n_{t+1}\times \mathchorus{RU}^{Pr}|   	
\end{gather} 

\section{Operational Design and Complexity}
The operational summary of REE-TM is depicted in Algorithm \ref{algo-Ree-tm}.   
REE-TM initializes a list of job requests ($List_{{\mathds{JR}}}$), list of virtual nodes ($List_{\mathds{VN}}$), and list of physical nodes ($List_{\mathds{PN}}$), where the heterogeneous internet service consumption environment  submits job requests \{$\lambda_1$, $\lambda_2$, ..., $\lambda_m$\} for  execution within cloud infrastructure.  For each time-interval \{$t$\}, heterogeneous workload is received and categorized on the basis of their resource usage and input/output demand depending on the interdependency among different jobs which belong to common user (Step 3). The jobs are allocated to available physical machines after computing and incorporating reliability (Steps 4 and 5). To estimate workload and traffic states in the next interval steps 6 and 7 are executed.     
\begin{figure}[!htbp]
	\removelatexerror
	\begin{algorithm}[H]
		\caption{REE-TM Operational Summary}
		\label{algo-Ree-tm}
		Initialize: $List_{{\mathds{JR}}}$, $List_{\mathds{VN}}$, $List_{\mathds{PN}}$\; 
		
		\For {each time-interval $\{t\}$}{ 
			
			{   Receive actual workload as $List_{{\mathds{JR}}}=$ \{$\lambda_1$, $\lambda_2$, ..., $\lambda_m$\} and categorize into $\mathds{C}^{St}$, $\mathds{C}^{Rh}$, and $\mathds{C}^{Nt}$ using Eq. (\ref{trafficstates}) \;  
				Allocate job requests on available physical nodes by applying Eqs. (\ref{re1}-\ref{cs}) \;
				Compute reliability of each $\mathchorus{PN}$ using concepts of Section \ref{rem}  before request allocation \;
				Estimate workload: CALL TG-QNN () to generate predicted traffic ($\mathchorus{TW}_{t+1}^{Pr}$) for next time-interval \;
				Compute workload entropy using concepts mentioned in Section \ref{tscea} for ${t+1}^{th}$ interval \;								
				
			}
			
		}

	\end{algorithm}
	
\end{figure}
\\ \textit{Time Complexity}:
Step 1 initializes lists of jobs for different users with a time complexity of $\mathcal{O}(1)$. Steps 3-7 iterate for $t$ time slots: Step 3 receives workload and categorizes traffic with a time complexity of $\mathcal{O}(1)$, and steps 4 and 5 also maintain a complexity of $\mathcal{O}(1)$. Step 6 employs the TG-QNN for predicting resource utilization, which has a complexity of $\mathcal{O}({Z} \times n^2 \times {M} \times {K})$ as detailed in Section \ref{Qdpbho}. Therefore, the overall computational complexity is $\mathcal{O}(t \times {Z} \times n^2 \times {M} \times {K})$.

\section{Performance Evaluation}

\subsection{Implementation}
A REE-TM prototype is configured with the collaboration of major modules discussed below:
\begin{itemize}
	\item \textit{Workload Estimation}: The future resource usage of different job requests is predicted by implementing TG-QNN () module which is trained with QBHO learning Algorithm \ref{algo-dpbho}. The workload is predicted to arrange physical resources proactively and error is evaluated.   
	\item \textit{Entropy Computation}: This module computes difference of actual and previous predicted workload, estimate various traffic states and use them to compute entropy for the upcoming workload. 
	\item \textit{User's Jobs submission}: Different users from internet service consumption environment   submit different job requests along with their priority of execution, to be assigned to virtual nodes which are allocated to available physical nodes. 
	
	\item \textit{Traffic State Categorization}: The submitted job requests are categorized according to various resource demands using Eq. (\ref{trafficstates}) and assorted into clusters of stragglers, resource hogs, and normal tasks. 
	\item \textit{Reliability Computation}: For each physical node, reliability is computed using concepts of Section \ref{reetm}. In this module MTBF is computed on the basis of number of failures observed in previous time interval.  	
	\item \textit{Jobs Allocation}: The job requests are allocated and executed on selected virtual node as per their resource requirement and virtual nodes are allocated to selected physical nodes by applying Eqs. (\ref{re1}-\ref{cs}).
	\item \textit{Evaluation}: The performance of REE-TM governed traffic management is measured by implementing computation of total reliability, energy-efficiency and resource utilization (Section \ref{reetm}), computing number of failures and success of load distribution observed during periodic time-intervals.
	
\end{itemize}

\subsection{Experimental Set-up and {Dataset}}
The simulation experiments are executed on a server machine assembled with two Intel\textsuperscript{\textregistered} Xeon\textsuperscript{\textregistered} Silver 4114 CPU with 40 core processor and 2.20 GHz clock speed. The server machine is deployed with 64-bit Ubuntu 16.04 LTS, having main memory of 128 GB. The data centre environment included three different types of servers and four types of virtual nodes configuration shown in Tables \ref{table:server} and \ref{table:vm} in Python. The resource features like power consumption ($PW_{max}, PW_{min}$), MIPS, RAM and memory are taken from real server IBM \cite{IBM1999} configuration where $S_1$ is `ProLiantM110G5XEON3075', $S_2$ is `IBMX3250Xeonx3480' and $S_3$ is `IBM3550Xeonx5675'. The virtual nodes configuration is inspired from the virtual machine instances of Amazon website \cite{amazon1999EC2}. {TG-QNN model was evaluated through classical simulation by implementing the mathematical equations governing qubit vector processing as defined in Eqs. (\ref{qubit})-(\ref{o3}) in Section IV. These equations simulate the Toffoli gate's effect and activation operations at hidden and output layers, enabling non-linear transformation of qubit states using deterministic matrix-based computations.} The experimental values or range of values for various intended parameters for TG-QNN optimization and job requests classification are mentioned in Table \ref{tab:qnn_params} and Table \ref{tab:task _classification}, respectively.

\begin{table}[!htbp]
	\centering
	
	\caption[Table caption text] {Physical Node Configuration}  
	\label{table:server}
	\resizebox{8.5cm}{!}{
		\begin{tabular}{lccccc}
			\hline
			Physical nodes&PE&MIPS&RAM(GB)&$PW_{max}$&$PW_{min}$/$PW_{idle}$\\
			\hline
			$S_1$ 	& 2&2660&4&135&93.7 \\
			$S_2$	& 4&3067&8&113&42.3 \\
			$S_3$	& 12&3067&16&222&58.4 \\

			\hline
	\end{tabular}}
\end{table}

\begin{table}[!htbp]
	\centering
	
	\caption[Table caption text] {Virtual Node Configuration}  
	\label{table:vm}
	\begin{tabular}{lccc}
		\hline
		Virtual node type& PE &MIPS&RAM(GB)\\
		\hline
		$\mathchorus{VN}_{1}$&1&500&0.5\\
		$\mathchorus{VN}_{2}$&2&1000&1\\
		$\mathchorus{VN}_{3}$&3&1500&2\\
		$\mathchorus{VN}_{4}$&4&2000&3\\

		\hline
	\end{tabular}
\end{table}

\begin{table}[h!]
\centering
\caption{Experimental parameters  for TG-QNN optimization}
\resizebox{0.49\textwidth}{!}{
\begin{tabular}{|l|p{4.2cm}|}
\hline
\textbf{Parameters}    & \textbf{Value/Range}                             \\ \hline
{Number of Qubit-Neurons}     & Input: 2-4, Hidden: 2-6, Output: 1             \\ \hline
{Toffoli Gate Configuration}  & Standard 3-qubit gate (2 controls, 1 target)     \\ \hline
{Initial Qubit State}         & Randomized (|0⟩ + |1⟩)/$\sqrt{2}$, (|0⟩ + i|1⟩)/$\sqrt{2}$  \\ \hline
{Learning Algorithm}          & Quantum Blackhole optimization               \\ \hline
{Population Size}             & 36-48                                            \\ \hline
{Event Horizon Radius}        & 0.05-0.1                                         \\ \hline
{Maximum Iterations}          & 1000-5000                                        \\ \hline
{Fitness Function}            & Fidelity, Mean Squared Error (MSE)               \\ \hline
{Crossover Probability}       & 0.7-0.9                                          \\ \hline
{Mutation Rate}               & 0.01-0.05                                        \\ \hline
{Quantum Noise}               & Low ($<$ 0.01)                                      \\ \hline
{Convergence Criteria}        & Fitness threshold or max iterations              \\ \hline
{Learning Rate}               & 0.001-0.01                                       \\ \hline
\end{tabular}}
\label{tab:qnn_params}
\end{table}

\begin{table}[h!]
\centering
\caption{Experimental parameters  for Job-request classification}
\resizebox{0.49\textwidth}{!}{
\begin{tabular}{|l|p{1.2cm}|p{1.2cm}|p{1.4cm}|}
\hline

\textbf{Status} &$\mathchorus{C}^{thr}$ & $\mathchorus{M}^{thr}$ & \textbf{Constraint}                                        \\ \hline

\textit{Straggler} ($St$) & 5\%-80\% & 80\%-99\% & $\mathchorus{C}^{thr} \bigwedge \mathchorus{M}^{thr}$
\\ \hline
\textit{Res. hogs} ($Rh$) & 80\%-99\% & 80\%-99\% & $\mathchorus{C}^{thr} \bigvee \mathchorus{M}^{thr}$
\\ \hline
\textit{Normal} ($Nt$) & 5\%-80\% & 5\%-80\% & $\mathchorus{C}^{thr} \bigwedge \mathchorus{M}^{thr}$
\\ \hline
\end{tabular}}
\label{tab:task _classification}
\end{table}

Google Compute Cluster  dataset (GCD) is utilized for performance estimation of REE-TM and comparative approaches which contains resources viz., CPU, memory, disk I/O  request and resource usage information of 672,300 jobs executed on 12,500 servers for the period of 29 days \cite{reiss2011google}. The CPU and memory utilization  percentage of VMs are obtained from the given CPU and memory usage percentage for each task in every five minutes over period of twenty-four hours. Table \ref{tab:ds} shows the statistical characteristics of the evaluated workloads. 
\begin{table}[!htbp]
	\caption{Characteristics of evaluated workloads}\label{tab:ds}
	\centering
	\resizebox{1.0\columnwidth}{!}{
		\begin{tabular}{lccccccc}
			\hline
			\multirow{2}{*}{Workload} & \multirow{2}{*}{Duration}  &No. of & Machines & Traces/ & \multirow{2}{*}{Mean} & \multirow{2}{*}{Std.Dev.} \\ 
			&& Machines & Selected & Machine && \\
			\hline
			GCD-CPU & 29 days & 125K & 1000 & 100K & 21.84& 13.62 \\ \hline
			GCD-Mem & 29 days & 125K & 1000 & 100K & 19.55& 16.6 \\ \hline
			
	\end{tabular}}
\end{table}

\subsection{Results}
Table \ref{table:performanceGCD} reports the performance metrics for REE-TM over continuous duration of 200 minutes:  average reliability ($\mathds{R}$ (\%)), average resource utilization ($\mathchorus{RU}$  (\%)), energy consumption ($E$ (KWH)), workload estimation error ({$MSE$}), entropy ($\mathds{E}n$), average number of overloads ($\mathds{OV}$ (\%)), number of active physical nodes (\textit{Active} $\mathchorus{PN}$), load allocation success rate ($\mathds{SUCC}$). The achieved reliability values lies in the range [85.6\%-97.8\%] which varies with the increasing number of job requests and availability of physical resources. Resource utilization lies in the range [42.6\% -49.9\%] independent of the number of job requests. The energy consumption increases with increasing number of job requests and number of active physical nodes. The workload estimation errors and traffic entropy varies according to the efficiency of TG-QNN prediction model. The number of failures or overloads lies in the range [1.95\%-8.20\%] and corresponding load distribution success rate is achieved up to 99.9\%. The number of active physical nodes i.e., 9 nodes  are  utilized to execute 100 job requests has reduced  up to 29\% i.e., 70 nodes are used for execution of 1000 job requests.

\begin{table*}[!htbp]
	
	\caption[Table caption text] {Performance metrics for REE-TM}  
	\label{table:performanceGCD}
	\small
	\centering 
	\resizebox{15cm}{!}{
		\centering
		\begin{tabular}{|l|c|c|c|c|c|c|c|c|c|}
			\hline
			
			$Request (\lambda)\# $&$T(min.)$& $\mathds{R}$ (\%) &$\mathchorus{RU}$ (\%)&$E$ (KWH) &$MSE$ &$\mathds{E}n$ &$\mathds{OV}$ (\%) &\textit{Active} $ \mathchorus{PN}$ &$ \mathds{SUCC}$  \\ \hline \hline			
			\multirow{3}{*}{100}
			&10& 96.6&42.6  & 1.015& 0.0008& 0.004 & 4.20 &9 &95.8 \\ \cline{2-10}
			&100&96.9 & 43.8 &0.919& 0.0004& 0.007&2.30&8 &97.7\\ \cline{2-10}
			&200& 97.8& 42.8&0.992&0.0003& 0.003&1.20&9 &98.8\\ \hline \hline
			
			\multirow{3}{*}{200}
			&10& 97.6&47.7  & 1.993& 0.0006& 0.008 & 1.30 &13 &98.7 \\ \cline{2-10}
			&100&97.9 & 47.5 &1.994& 0.0016& 0.003&3.40&18 &96.6\\ \cline{2-10}
			&200& 97.8& 47.9&1.993&0.0009& 0.006&5.20&17 &94.8\\ \hline \hline

			\multirow{3}{*}{400}
			&10& 85.9&49.6  & 4.041& 0.0004& 0.008 & 6.11 &30 &94.0 \\ \cline{2-10}
			&100&85.9 & 49.8 &4.019& 0.0004& 0.009&5.63&28 &97.7\\ \cline{2-10}
			&200& 86.6& 49.9&4.019&0.0006& 0.005&3.80&29 &96.2\\ \hline \hline
			
			\multirow{3}{*}{600}
			&10& 86.6&48.9  & 6.095& 0.0002& 0.005 & 4.11 &39 &95.9 \\ \cline{2-10}
			&100&86.9 & 48.9 &6.019& 0.0006& 0.019&2.30&38 &97.7\\ \cline{2-10}
			&200& 87.8& 48.8&6.192&0.0003& 0.006&1.80&41 &98.2\\ \hline \hline
			
			\multirow{3}{*}{800}
			&10& 86.3&42.9  & 8.357& 0.0005& 0.004 & 4.12&60 &95.9 \\ \cline{2-10}
			&100&87.9 & 43.1 &8.358& 0.0004& 0.005&1.29&58 &98.7\\ \cline{2-10}
			&200& 86.4& 42.6&8.361&0.0008& 0.003&0.03&59 &99.8\\ \hline \hline
			
			\multirow{3}{*}{1000}
			&10& 85.6&48.6  & 10.015& 0.0005& 0.008 & 6.20 &69 &96.2 \\ \cline{2-10}
			&100&85.9 & 48.8 &10.019& 0.0005& 0.005&4.03&70 &97.7\\ \cline{2-10}
			&200& 86.8& 48.9&10.092&0.0004& 0.007&7.05&68 &98.1\\ \hline 
	\end{tabular}}
\end{table*}

\par The prediction accuracy of TG-QNN is evaluated under two scenarios: (1) varying prediction intervals (window sizes) and (2) varying job request volumes. For scenario (1), GCD workload traces are aggregated into intervals of 5 to 60 minutes and normalized to the range [0,1] using Eq.~(\ref{normalisation}), with $x_a = 0.0001$ and $x_b = 0.999$. 

\begin{equation}
\label{normalisation}
\hat{D^{In}} = x_a + \frac{d_i - D^{In}_{min}}{D^{In}_{max} - D^{In}_{min}} \times x_b
\end{equation}

\begin{figure}[!htbp]
	\centering	
	\subfigure[ GCD Memory usage]{\includegraphics[width=0.4\linewidth, scale=2]{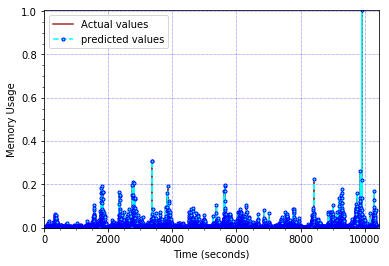}} 
	\subfigure[ GCD CPU usage]{\includegraphics[width=0.4\linewidth, scale=2]{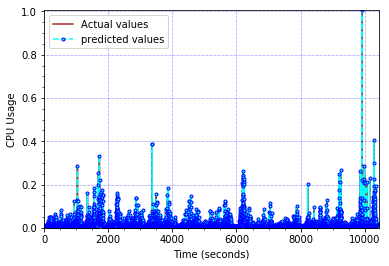}}
	\caption{ Actual v/s predicted service requests per second}
	\label{gcd_sec}	
\end{figure}

\begin{figure}[!htbp]
	\centering	
	\subfigure[ Memory usage]{\includegraphics[width=0.4\linewidth, scale=2]{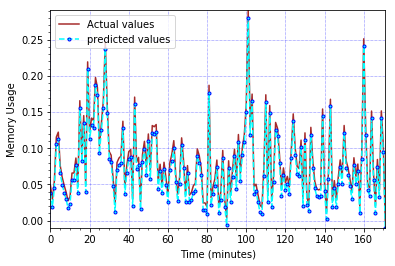}} 
	\subfigure[ CPU usage]{\includegraphics[width=0.4\linewidth, scale=2]{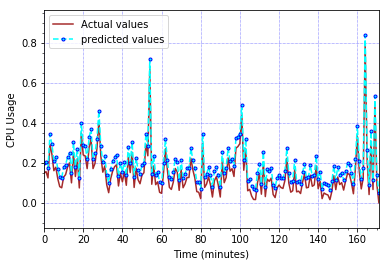}}
	\caption{  Actual v/s predicted service requests per minute}
	\label{gcd_min}	
\end{figure}

Fig.~\ref{gcd_sec} presents the predicted versus actual memory and CPU usage for 10,000 samples at 1-minute intervals, showing near-perfect overlap with average prediction errors of 0.0005 and 0.0006, respectively. Fig.~\ref{gcd_min} illustrates results for 1-hour intervals (160+ samples), with slightly higher average errors of 0.004 (memory) and 0.006 (CPU). These results confirm that TG-QNN achieves high prediction accuracy, particularly for shorter intervals, due to denser sample availability.

\begin{figure}[!htbp]
	\centering	
	\subfigure[  Memory usage]{\includegraphics[width=0.4\linewidth, scale=2]{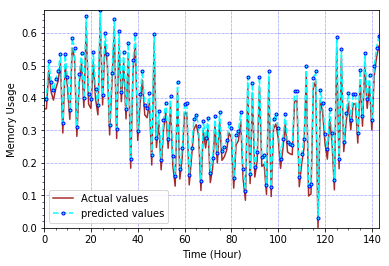}} 
	\subfigure[  CPU usage]{\includegraphics[width=0.4\linewidth, scale=2]{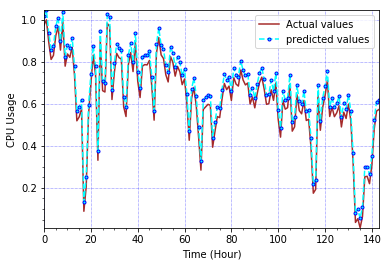}}
	\caption{Actual v/s predicted service requests for a random virtual node}
	\label{gcd_vm}	
\end{figure}

For the \textit{second scenario}, compute and storage usage data for job requests extracted from Google compute cluster traces are sampled at 5-minute intervals. These CPU and memory usage values for different virtual nodes are normalized and used for training the workload prediction model. Fig.~\ref{gcd_vm} shows the comparison of actual and predicted resource usage for a randomly selected virtual node, where predicted CPU (Fig.~\ref{gcd_vm}~(a)) and memory (Fig.~\ref{gcd_vm}~(b)) values closely align with actual usage. The average prediction errors are 0.00344 (CPU) and 0.00091 (memory), demonstrating TG-QNN's high accuracy in modeling dynamic workloads, enabled by its quantum-based learning mechanism and adaptive weight tuning via the Quantum Blackhole Learning algorithm.

\begin{figure}[!htbp]
	\centering
	\includegraphics[width=0.7\linewidth]{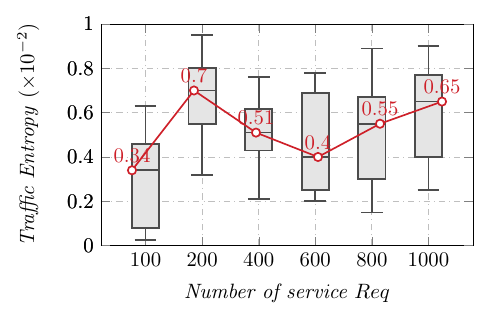}
	\caption{{Entropy Analysis}}
	\label{fig:gcdentropy}
\end{figure}

\begin{figure}[!htbp]
	\centering
	\includegraphics[width=0.65\linewidth]{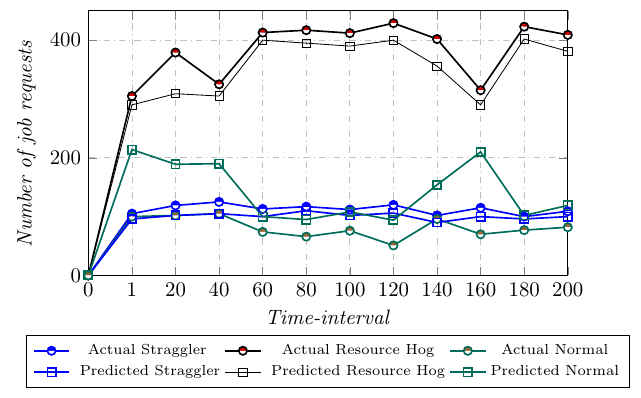}
	\caption{Actual v/s Predicted Traffic states: Straggler,  Hogs, and Normal}
	\label{fig:gcdstates}
\end{figure}

The range of entropy values achieved during workload estimation for various number of job requests is reported via box-plots in Fig. \ref{fig:gcdentropy}. The entropy values are computed using Eqs. (\ref{entropy1} and \ref{entropy2}) for duration of 200 minutes for varying number of requests. The median values for different number of job requests lies in the range [0.0003-0.00075] while the values for lower quartile and upper quartile  lie in the ranges  [0.0001-0.0004] and [0.0006-0.001], respectively. The values of entropy varies according to the deviation expected in the upcoming workload. Fig. \ref{fig:gcdstates} presents  actual versus predicted traffic states: Straggler, Resource Hogs, and Normal over continuous time period for 600 job requests.

\subsection{Comparison and Discussion}

The proactive workload estimation efficiency of TG-QNN  for upcoming job requests composing internet traffic is compared with the graph neural network (GNN) \cite{li2024evogwp}, attention-based method (ATTN) \cite{zhao2024tfegru}, Evolutionary Quantum Neural Network (EQNN) \cite{singh2021quantum}, and the Evolutionary Neural Network (ENN) \cite{saxena2021op}. Since no existing traffic management approach leverages traffic categorization-based reliable load distribution, direct experimental comparison is not applicable. However, the proposed approach is evaluated against three scenarios: \textit{OPTIMAL}, \textit{W-REE-TM$^\ast$}, and \textit{W-REE-TM$^{\ast\ast}$}.

\begin{itemize}
    \item \textit{OPTIMAL:} Assumes perfect foresight resource usage, traffic categorization, and node reliability are known without prediction error or deviation. This provides the theoretical best-case baseline for performance metrics.
    
    \item \textit{W-REE-TM$^\ast$:} A baseline without workload prediction (i.e., REE-TM without TG-QNN), but includes reliability-aware load allocation. Traffic is assigned without prior traffic state knowledge, though node reliability is still considered.
    
    \item \textit{W-REE-TM$^{\ast\ast}$:} A basic traffic management variant excluding both workload prediction and reliability integration. Traffic is allocated with fixed virtual node configurations and no reliability assessment.
\end{itemize}

\subsubsection{Workload Analysis}

{Table~\ref{comp} compares the proposed TG-QNN model with state-of-the-art methods: GNN~\cite{li2024evogwp}, ATTN~\cite{zhao2024tfegru}, EQNN~\cite{singh2021quantum}, and ENN~\cite{saxena2021op} across three prediction window sizes (PWS): 5, 10, and 60 minutes, using MSE and MAE as evaluation metrics. TG-QNN achieves the best short-term accuracy at PWS = 5, with an MSE of 0.0005 and MAE of 0.0099, outperforming even quantum- and attention-based models. While ATTN slightly outperforms TG-QNN at PWS = 10, TG-QNN remains highly competitive. At PWS = 60, TG-QNN again leads with the lowest MSE (0.0024) and MAE (0.0336). These results highlight TG-QNN’s robust performance across varying time horizons, demonstrating its scalability and effectiveness for real-time workload forecasting in dynamic cloud environments.
}
\begin{table}[!htbp] 	
	\caption[Table caption text] {{Workload analysis: TG-QNN v/s state-of-the-arts} }  
	\label{comp} 	
	\centering
	\resizebox{8.7cm}{!}{
		\begin{tabular}{  |l|  l | l | l |  l|  l| l| }
			\hline
			\textbf{PWS}	&\textbf{Error} &\textbf{ GNN \cite{li2024evogwp}} & \textbf{ATTN \cite{zhao2024tfegru}}& \textbf{EQNN \cite{singh2021quantum}} & \textbf{ENN \cite{saxena2021op}} & \textbf{TG-QNN}\\
			
			\hline
			\multirow{2}{*}{\textbf{5 min }} & MSE &0.0533 & 0.0022 & 0.0021 &0.0032 & \textbf{0.0005}\\  \cline{2-7}
            
			& MAE &0.1178 &0.0323 & 0.0334&0.0666 &\textbf{0.0099}\\
		
			\hline

            \multirow{2}{*}{\textbf{10 min }} &MSE &0.0445 &\textbf{0.0022} &0.0025 & 0.0031 & 0.0024\\  \cline{2-7}
            
			&MAE  & 0.5457 & \textbf{0.0330}& 0.0431& 0.0661& 0.0342\\
		
			\hline

        \multirow{2}{*}{\textbf{60 min }} &MSE & 0.0486 &0.0027 & 0.0052&0.0093 &\textbf{0.0024} \\  \cline{2-7}
            
			&MAE  & 0.5304&0.0356 & 0.0872 & 0.0993&\textbf{0.0336}\\
		
			\hline
	\end{tabular}}
  \\ \footnotesize{ PWS : Prediction Window Size, GNN: Graph Neural Network, ATTN: Attention-based method, EQNN: Evolutionary Quantum Neural Network}
\end{table}

\begin{figure}[!htbp]
	\centering
	\includegraphics[width=0.7\linewidth]{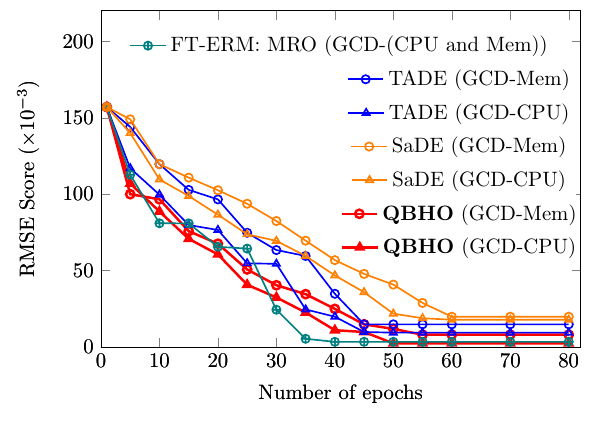}
	\caption{Convergence analysis: RMSE over consecutive epochs}
	\label{fig:convergence}
\end{figure}

{The convergence graph depicted in Fig. \ref{fig:convergence} reveals that QBHO consistently achieves faster and more stable convergence in both the datasets: GCD-CPU and GCD-Mem settings, outperforming TADE, SaDE, and FT-ERM. Its rapid RMSE reduction and early saturation highlight the effectiveness of its quantum-inspired event horizon mechanism in accelerating global search and avoiding premature convergence.}

{Table~\ref{tab:complexity} presents a comparative analysis of four optimization algorithms based on their computational complexity, space requirements, and average classification accuracy. Although all methods exhibit similar time complexity on the order of $\mathcal{O}(Zn^2MK)$, QBHO clearly demonstrates superiority by achieving highest average accuracy (98.06\%) with lowest space complexity of $\mathcal{O}(Zn^2)$. In contrast, Self-adaptive Differential Evolution (SaDE) optimization \cite{singh2021quantum} and Tri-adaptive Differential Evolution (TADE) \cite{saxena2020proactive} require twice the memory with lower accuracy, and although Multiple Resource Optimization (MRO) \cite{saxena2022fault} incurs higher computational cost $\mathcal{O}(2ZMKn^2)$, it delivers the lowest accuracy. These results emphasize optimal trade-off between efficiency and predictive performance of QBHO, making it an effective choice for applications with memory constraints and accuracy sensitive.}

\begin{table}[h!]
\centering
\caption{{Computational complexity and average accuracy analysis}}
\resizebox{0.45\textwidth}{!}{
\begin{tabular}{|p{2.2cm}|p{3.0cm}|p{2.2cm}|p{2.0cm}|}
\hline

\textbf{Optimization} & \textbf{Time Complexity}& \textbf{Space Complexity} &  \textbf{Avg. accuracy}                                       \\ \hline

SaDE \cite{singh2021quantum} &  $\mathcal{O}(Zn^2MK)$ &$\mathcal{O}(2Zn^2)$  & 97.76\%   \\ \hline
TADE \cite{saxena2020proactive} & $\mathcal{O}(Zn^2MK)$  &$\mathcal{O}(2Zn^2)$ & 95.75\%   \\ \hline
MRO \cite{saxena2022fault} &  $\mathcal{O}(2ZMKn^2)$  & $\mathcal{O}(4Zn^2)$ & 93.04\% \\ \hline
QBHO  & $\mathcal{O}(Zn^2MK)$  & $\mathcal{O}(Zn^2)$  & 98.06\%\\ \hline

\end{tabular}}
\\ \footnotesize{ $Z$: number of vectors or population size, $n$: number of nodes in input layer, $M$: number of samples, $K$: number of iterations}
\label{tab:complexity}
\end{table}


\subsubsection{Traffic Management}
\begin{figure}[!htbp]
	\centering
	\includegraphics[width=0.6\linewidth]{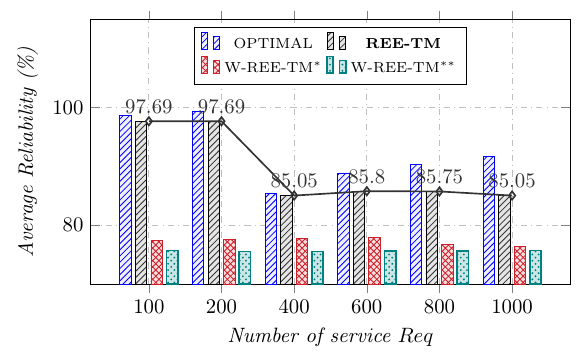}
	\caption{Reliability}
	\label{fig:gcdreliability}
\end{figure}

\paragraph{Reliability}
The average reliability of REE-TM is compared with Optimal case, W-REE-TM$^{\ast}$, and W-REE-TM$^{\ast\ast}$ in Fig. \ref{fig:gcdreliability}. The reliability is computed for each physical node before allocation of workload with respect to resource availability and probable occurrence of job failure. The workload is assigned to a physical node only if it satisfies a minimum reliability criteria and the average reliability is computed for the aggregate of the total number of active physical nodes for the entire duration of the job request execution. 
The average reliability of REE-TM remains close to OPTIMAL, starting at 97\% for 100–200 job requests and decreasing to 85\% as virtual node migrations occur due to resource limitations over 200 minutes. REE-TM’s reliability is only 1–6\% lower than OPTIMAL, primarily due to workload and entropy estimation errors from TG-QNN. Nonetheless, REE-TM achieves significantly higher reliability up to 26.04\% and 30.25\% compared to W-REE-TM$^{\ast}$ and W-REE-TM$^{\ast\ast}$, respectively. The absence of proactive workload estimation in these baseline approaches results in higher migration rates and reduced reliability. W-REE-TM$^{\ast}$ performs better than W-REE-TM$^{\ast\ast}$ due to its one-time reliability computation before allocation; however, static reliability assessment fails to adapt to dynamic resource changes during execution, leading to further migrations and reliability loss.

\begin{figure}[!htbp]
	\centering
	\includegraphics[width=0.6\linewidth]{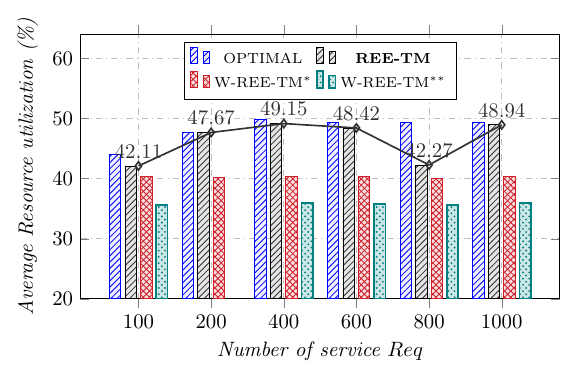}
	\caption{Resource utilization}
	\label{fig:gcdru}
\end{figure}

\paragraph{Resource utilization}
Fig. \ref{fig:gcdru} entails the comparison of overall resource  utilization of REE-TM with OPTIMAL, W-REE-TM$^{\ast}$ and W-REE-TM$^{\ast\ast}$ for varying number of requests. REE-TM achieved an average resource utilization in the range [42\%-49\%] which is improved up to 22.35\% and 39.81\% over W-REE-TM$^{\ast}$ and W-REE-TM$^{\ast\ast}$, respectively. Owing to the fact that the resource utilization capability depends upon the adopted strategy for workload allocation, REE-TM provides effective reliability at the cost of [1\%-4\%] lesser resource utilization as compared with the optimal case.

\begin{figure}[!htbp]
	\centering
	\includegraphics[width=0.6\linewidth]{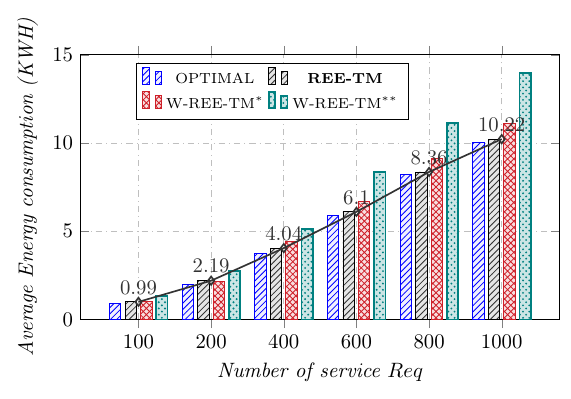}
	\caption{Energy Consumption}
	\label{fig:gcdenergy}
\end{figure}

\paragraph{Energy Consumption}
Fig.~\ref{fig:gcdenergy} compares energy consumption across varying service request volumes. While REE-TM consumes 1--5\% more energy than OPTIMAL due to minor workload estimation errors, it achieves a 20--23\% reduction compared to W-REE-TM$^{\ast}$ and W-REE-TM$^{\ast\ast}$. The lack of proactive workload estimation in the latter approaches leads to overprovisioning via high-capacity virtual nodes, activating more physical nodes and increasing energy use. In contrast, REE-TM leverages TG-QNN-based estimation to minimize active nodes and improve energy efficiency.

\begin{figure}[!htbp]
	\centering
	\includegraphics[width=0.6\linewidth]{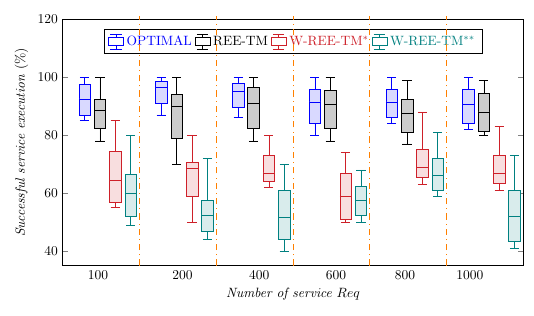}
	\caption{Load execution success}
	\label{fig:gcdsuccess-rate}
\end{figure}

\paragraph{Load Distribution Analysis}
The workload distribution performance of REE-TM is evaluated against OPTIMAL, W-REE-TM$^{\ast}$, and W-REE-TM$^{\ast\ast}$ in terms of successful service execution, failure rate, and number of active physical nodes for varying job requests. As shown in Fig.~\ref{fig:gcdsuccess-rate}, REE-TM achieves a success rate close to OPTIMAL, with upper quartiles reaching 100\% for up to 600 requests and trailing OPTIMAL by only 1–2\% beyond that. Median success rates of REE-TM remain nearly identical to OPTIMAL across all scenarios. REE-TM outperforms W-REE-TM$^{\ast}$ and W-REE-TM$^{\ast\ast}$ by up to 38\% and 74\%, respectively. Additionally, W-REE-TM$^{\ast}$ exceeds W-REE-TM$^{\ast\ast}$ by 30\% due to the inclusion of reliable job assignment.

\begin{figure}[!htbp]
	\centering
	\includegraphics[width=0.6\linewidth]{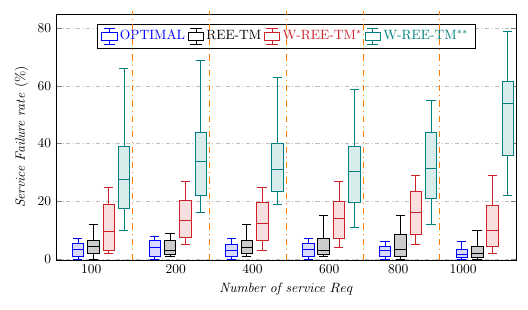}
	\caption{Load execution failure}
	\label{fig:gcdfailurerate}
\end{figure}

Fig.~\ref{fig:gcdfailurerate} presents the box-plot comparison of job failures over 200 minutes across varying service requests. Failure rates are lowest for OPTIMAL and REE-TM, and highest for W-REE-TM$^{\ast\ast}$. For 1000 requests, REE-TM reduces failure occurrences by up to 66.67\% and 87.9\% compared to W-REE-TM$^{\ast}$ and W-REE-TM$^{\ast\ast}$, respectively, based on upper quartile values. Median and lower quartile failure rates for REE-TM align with OPTIMAL, while slightly higher values are observed in the upper quartile. This performance gain stems from REE-TM’s accurate workload estimation, effective traffic-state categorization, and reliability-aware job execution.

\begin{figure}[!htbp]
	\centering
	\includegraphics[width=0.6\linewidth]{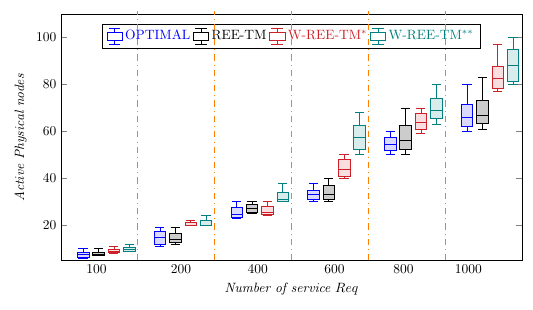}
	\caption{Active physical nodes}
	\label{fig:gcdapms}
\end{figure}

Fig.~\ref{fig:gcdapms} illustrates the number of active physical nodes across varying job request volumes for four approaches. As expected, active nodes increase with workload; however, the scaling trend consistently follows: W-REE-TM$^{\ast\ast}$ $>$ W-REE-TM$^{\ast}$ $>$ REE-TM $\geq$ OPTIMAL. Notably, REE-TM reduces active nodes by up to 11.11\% and 15.79\% compared to W-REE-TM$^{\ast}$ and W-REE-TM$^{\ast\ast}$, respectively, as indicated by the upper quartile of the box plots for 1000 job requests. This reduction is enabled by REE-TM's effective traffic-state-based workload categorization, allowing resource-aware selection of the most suitable physical nodes, thereby maximizing consolidation and minimizing active node count.

Table \ref{summaryComparison} presents a comparison of values obtained for different parameters of REE-TM with relative values presented in state-of-the-art approaches \cite{saxena2021op}, \cite{saxena2022fault}, \cite{kaur2021energy}, \cite{saxena2022high}, and  \cite{kaur2019keids} which have used Google compute cluster data traces for performance evaluation.

{Table \ref{summaryComparison} compares the performance of the proposed REE-TM method with state-of-the-art approaches. REE-TM demonstrates competitive workload estimation accuracy (96\%-99.5\%), reliability (up to 95.7\%), and success rate (91\%-100\%), making it comparable to leading methods like, OP-MLB and SRE-HM. A key contribution of REE-TM is its superior energy efficiency (10.092 KWH), which is notably lower  electrical power consumption than methods such as OP-MLB (3.59E+04 W) and FT-ERM (37.67 KW), highlighting its potential for sustainable resource management. Although its resource utilization (49.7\%) is lower than SRE-HM (74.9\%), REE-TM offers a balanced and reliable solution, excelling in reliability and energy efficiency, which position it as a strong contender for real-world applications.}
\begin{table*}[!h]
	
	\caption{{Performance parameter comparison: REE-TM v/s State-of-the-arts}}
	\label{summaryComparison}
	\scriptsize
	\centering
	\resizebox{0.80\textwidth}{!}{
		\centering
		\begin{tabular}{|p{1.8cm}|| p{1.7cm} |p{1.5cm} |p{1.5cm}|p{1.5cm}|p{1.6cm}| p{1.6cm}| }
			\hline

			Approach&{\textit{Workload Estimation Accuracy (\%) }}&{\textit{Reliability (\%)}}&{\textit{Resource Utilization (\%)}}& {\textit{Energy/Power Consumption }}& {\textit{Success Rate (\%)}}& {\textit{Failure Rate (\%) }}\\ \hline \hline
			OP-MLB \cite{saxena2021op}& 97.8\% &  $\times$ & 63.87 &3.59E+ 04 W & 94.48\% & 5.52\% \\ 	\hline
			FT-ERM \cite{saxena2022fault}&88\%-94\% & $\times$ & 63.03\% &37.67 KW &  95\%-99.8\% & 0.2\%-5.5\% \\ 	\hline
			En-SLA \cite{kaur2021energy}&  $\times$ &  $\times$ & varying & varying&  $\times$  &  $\times$ \\ 	\hline
			SRE-HM \cite{saxena2022high}& 90\%-98.6\% &$\times$&74.9\% &8578.4 W &  95\%-99.3\% & 0.7\%-4.62\% \\ 	\hline
			KEIDS \cite{kaur2019keids}& $\times$ &  $\times$ & CPU (67\%)&varying & $\times$ & $\times$ \\ 	\hline
			\textbf{REE-TM} & 96\%-99.5\% & up to 95.7\% & 49.7\% & 10.092 KWH & 91\%-100\% &0\%-9\%
			\\ 	\hline
			
	\end{tabular}}
\end{table*}

\subsubsection{Ablation Study}
{Table~\ref{tab:ablation} presents an ablation analysis of the proposed REE-TM framework over 1000 service requests, demonstrating the incremental contribution of key components: TGQNN, QBHO, and TSECE. The base model \textit{TGQNN+TM} achieves moderate performance with 66.67\% reliability ($\mathds{R}$), 80.98\% success rate ($\mathds{SUCC}$), and a low MSE of 0.0075. Integrating \textit{QBHO} (\textit{TGQNN+QBHO+TM}) significantly boosts reliability to 73.78\% and accuracy (MSE drops to 0.0003), reflecting the optimizer's effectiveness in refining load allocation. Adding the \textit{TSECE} mechanism (\textit{TGQNN+QBHO+TSECE+TM}) further elevates reliability to 86.88\% and success rate to 98.6\%, showcasing its role in enhancing execution consistency. The slight increase in energy consumption ($E$) remains within an acceptable range, indicating the trade-off is worthwhile for higher performance.} 

\begin{table}[h!]
\centering
\caption{{Ablation Analysis for 1000 service requests}}
\resizebox{0.45\textwidth}{!}{
\begin{tabular}{|p{3.5cm}|c|c|c|c|}
\hline

\textbf{REE-TM Versions} & $\mathds{R}$ (\%)& $MSE$ &   $\mathds{SUCC}$ (\%) & $E$ (KWH)                                    \\ \hline
TGQNN+TM & 66.67  & 0.0075& 80.98& 9.979 \\ \hline

TGQNN+QBHO+TM & 73.78 & 0.0003&88.98 & 10.119 \\ \hline
TGQNN+QBHO+TSECE+TM & 86.88 & 0.0004& 98.6& 10.092 \\ \hline

\end{tabular}}
\label{tab:ablation}
\end{table}
\section{Conclusion and Future Work}

This work proposed REE-TM, a reliability-focused approach for managing heterogeneous internet traffic in cloud computing environments. By categorizing incoming jobs into stragglers, resource hogs, and normal jobs based on resource usage and input/output requirements, REE-TM ensures optimal assignment to suitable virtual nodes. These nodes, hosted on physical servers with periodically assessed reliability, support maximum uptime during job execution. The TG-QNN prediction model optimized Quantum Blackhole learning algorithm, is developed to estimate the cloud traffic bursts and resource needs while calculating traffic entropy to manage congestion effectively. The  performance evaluation and comparison using a real-world benchmark data traces validates the capabilities of the proposed REE-TM approach in maximizing the reliability with effective resource utilization of cloud resources.    The future work will enhance REE-TM's scalability and adaptability, exploring advanced quantum optimization techniques and integrating diverse traffic patterns.




\bibliographystyle{IEEEtran} 
\bibliography{bibfile}

\begin{thebibliography}{10}
\providecommand{\url}[1]{#1}
\csname url@samestyle\endcsname
\providecommand{\newblock}{\relax}
\providecommand{\bibinfo}[2]{#2}
\providecommand{\BIBentrySTDinterwordspacing}{\spaceskip=0pt\relax}
\providecommand{\BIBentryALTinterwordstretchfactor}{4}
\providecommand{\BIBentryALTinterwordspacing}{\spaceskip=\fontdimen2\font plus
\BIBentryALTinterwordstretchfactor\fontdimen3\font minus
  \fontdimen4\font\relax}
\providecommand{\BIBforeignlanguage}[2]{{%
\expandafter\ifx\csname l@#1\endcsname\relax
\typeout{** WARNING: IEEEtran.bst: No hyphenation pattern has been}%
\typeout{** loaded for the language `#1'. Using the pattern for}%
\typeout{** the default language instead.}%
\else
\language=\csname l@#1\endcsname
\fi
#2}}
\providecommand{\BIBdecl}{\relax}
\BIBdecl

\bibitem{chakraborty2024optimizing}
T.~Chakraborty, C.~Kopp, and A.~N. Toosi, ``Optimizing renewable energy
  utilization in cloud data centers through dynamic overbooking: An mdp-based
  approach,'' \emph{IEEE Transactions on Cloud Computing}, 2024.

\bibitem{kaur2021energy}
K.~Kaur, S.~Garg, G.~Kaddoum, and N.~Kumar, ``Energy and sla-driven mapreduce
  job scheduling framework for cloud-based cyber-physical systems,'' \emph{ACM
  Transactions on Internet Technology (TOIT)}, vol.~21, no.~2, pp. 1--24, 2021.

\bibitem{al2019energy}
H.~M. Al-Kadhim and H.~S. Al-Raweshidy, ``Energy efficient and reliable
  transport of data in cloud-based iot,'' \emph{IEEE Access}, vol.~7, pp.
  64\,641--64\,650, 2019.

\bibitem{kang2021adaptive}
K.~Kang, D.~Ding, H.~Xie, Q.~Yin, and J.~Zeng, ``Adaptive drl-based task
  scheduling for energy-efficient cloud computing,'' \emph{IEEE Transactions on
  Network and Service Management}, vol.~19, no.~4, pp. 4948--4961, 2021.

\bibitem{gai2017sa}
K.~Gai, L.~Qiu, M.~Chen, H.~Zhao, and M.~Qiu, ``Sa-east: security-aware
  efficient data transmission for its in mobile heterogeneous cloud
  computing,'' \emph{ACM Transactions on Embedded Computing Systems (TECS)},
  vol.~16, no.~2, pp. 1--22, 2017.

\bibitem{cloudoutages2024}
\BIBentryALTinterwordspacing
T.~I. Group, ``2024 cloud service outages survey: The biggest failures and
  their impact,'' July 2024, available at Tech Insights Reports. [Online].
  Available:
  \url{https://www.techinsights.com/reports/2024-cloud-service-outages-survey}
\BIBentrySTDinterwordspacing

\bibitem{saxena2021op}
D.~Saxena, A.~K. Singh, and R.~Buyya, ``\uppercase{OP-MLB}: An online vm
  prediction based multi-objective load balancing framework for resource
  management at cloud datacenter,'' \emph{IEEE Transactions on Cloud
  Computing}, vol. doi: 10.1109/TCC.2021.3059096, 2021.

\bibitem{saxena2022high}
D.~Saxena and A.~K. Singh, ``A high availability management model based on vm
  significance ranking and resource estimation for cloud applications,''
  \emph{IEEE Trans. on Serv. Comput.}, 2022.

\bibitem{saxena2022fault}
D.~Saxena, I.~Gupta, A.~K. Singh, and C.-N. Lee, ``A fault tolerant elastic
  resource management framework towards high availability of cloud services,''
  \emph{IEEE Trans. on Netw. and Serv. Mgmt.}, 2022.

\bibitem{kaur2019keids}
K.~Kaur, S.~Garg, G.~Kaddoum, S.~H. Ahmed, and M.~Atiquzzaman, ``Keids:
  Kubernetes-based energy and interference driven scheduler for industrial iot
  in edge-cloud ecosystem,'' \emph{IEEE Internet of Things Journal}, vol.~7,
  no.~5, pp. 4228--4237, 2019.

\bibitem{ming2023adaptive}
F.~Ming, W.~Gong, and L.~Gao, ``Adaptive auxiliary task selection for
  multitasking-assisted constrained multi-objective optimization [feature],''
  \emph{IEEE Computational Intelligence Magazine}, vol.~18, no.~2, pp. 18--30,
  2023.

\bibitem{gai2017resource}
K.~Gai, M.~Qiu, H.~Zhao, and X.~Sun, ``Resource management in sustainable
  cyber-physical systems using heterogeneous cloud computing,'' \emph{IEEE
  Transactions on Sustainable Computing}, vol.~3, no.~2, pp. 60--72, 2017.

\bibitem{nesmachnow2015efficient}
S.~Nesmachnow, S.~Iturriaga, and B.~Dorronsoro, ``Efficient heuristics for
  profit optimization of virtual cloud brokers,'' \emph{IEEE computational
  intelligence magazine}, vol.~10, no.~1, pp. 33--43, 2015.

\bibitem{manogaran2022optimal}
G.~Manogaran, B.~S. Rawal, H.~Song, H.~Wang, C.~Hsu, V.~Saravanan, S.~N. Kadry,
  and P.~M. Shakeel, ``Optimal energy-centric resource allocation and
  offloading scheme for green internet of things using machine learning,''
  \emph{ACM Trans. on Internet Tech. (TOIT)}, vol.~22, no.~2, pp. 1--19, 2022.

\bibitem{muthanna2019secure}
A.~Muthanna, A.~A.~Ateya, A.~Khakimov, I.~Gudkova, A.~Abuarqoub, K.~Samouylov,
  and A.~Koucheryavy, ``Secure and reliable iot networks using fog computing
  with software-defined networking and blockchain,'' \emph{Journal of Sensor
  and Actuator Networks}, vol.~8, no.~1, p.~15, 2019.

\bibitem{zhao2024tfegru}
F.~Zhao, W.~Lin, S.~Lin, H.~Zhong, and K.~Li, ``Tfegru: Time-frequency enhanced
  gated recurrent unit with attention for cloud workload prediction,''
  \emph{IEEE Transactions on Services Computing}, 2024.

\bibitem{li2024evogwp}
J.~Li, J.~Yao, D.~Xiao, D.~Yang, and W.~Wu, ``Evogwp: Predicting long-term
  changes in cloud workloads using deep graph-evolution learning,'' \emph{IEEE
  Transactions on Parallel and Distributed Systems}, vol.~35, no.~3, pp.
  499--516, 2024.

\bibitem{singh2021quantum}
A.~K. Singh, D.~Saxena, J.~Kumar, and V.~Gupta, ``A quantum approach towards
  the adaptive prediction of cloud workloads,'' \emph{IEEE Transactions on
  Parallel and Distributed Systems}, 2021.

\bibitem{gupta2024multiple}
I.~Gupta, D.~Saxena, A.~K. Singh, and C.-N. Lee, ``A multiple controlled
  toffoli driven adaptive quantum neural network model for dynamic workload
  prediction in cloud environments,'' \emph{IEEE Transactions on Pattern
  Analysis and Machine Intelligence}, 2024.

\bibitem{minas2009energy}
L.~Minas and B.~Ellison, \emph{Energy efficiency for information technology:
  How to reduce power consumption in servers and data centers}.\hskip 1em plus
  0.5em minus 0.4em\relax Intel Press, 2009.

\bibitem{IBM1999}
IBM, ``Power model. [online].'' \emph{https:// www.ibm.com/}, 1999.

\bibitem{amazon1999EC2}
Amazon, ``Amazon ec2 instances. [online].'' \emph{https://
  aws.amazon.com/ec2/instance-types/}, 1999.

\bibitem{reiss2011google}
C.~Reiss, J.~Wilkes, and J.~L. Hellerstein, ``Google cluster-usage traces:
  format+ schema,'' \emph{Google Inc., White Paper}, pp. 1--14, 2011.

\bibitem{saxena2020proactive}
D.~Saxena and A.~K. Singh, ``A proactive autoscaling and energy-efficient vm
  allocation framework using online multi-resource neural network for cloud
  data center,'' \emph{Neurocomputing}, 2020.

\end{thebibliography}

\begin{IEEEbiography}[{\includegraphics[width=0.7\linewidth]{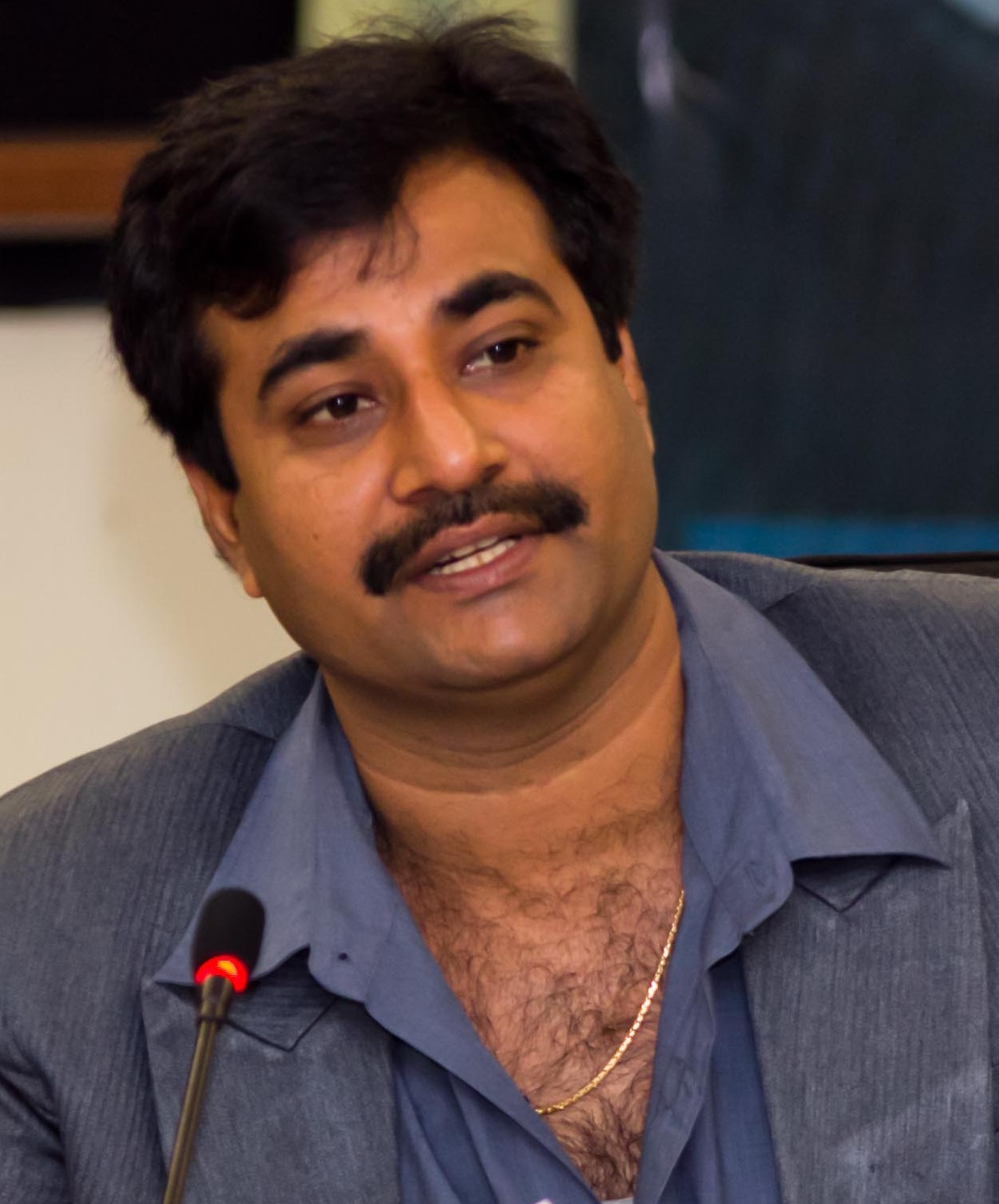}}]{Ashutosh Kumar Singh}is working as a Professor and Director of the Indian Institute of Information Technology Bhopal, India. Also, he is working as an Adjunct Professor in the VIZJA University, Warsaw, Poland. He received his Ph.D. in Electronics Engineering from Indian Institute of Technology, BHU, India and Post Doc from the Department of Computer Science, University of Bristol, UK. He has research and teaching experience in various Universities of India, the UK, and Malaysia.  He has published more than 400 research papers in different journals and conferences of high repute.  His research paper, published in the IEEE Transactions on Cloud Computing Journal, was honored with the 2022 Best Paper Award by the IEEE Computer Society Publications Board. His research area includes the Design and Testing of Digital Circuits, Data Science, Cloud Computing, Machine Learning, and Security.\end{IEEEbiography}
\begin{IEEEbiography}[{\includegraphics[width=0.9\linewidth]{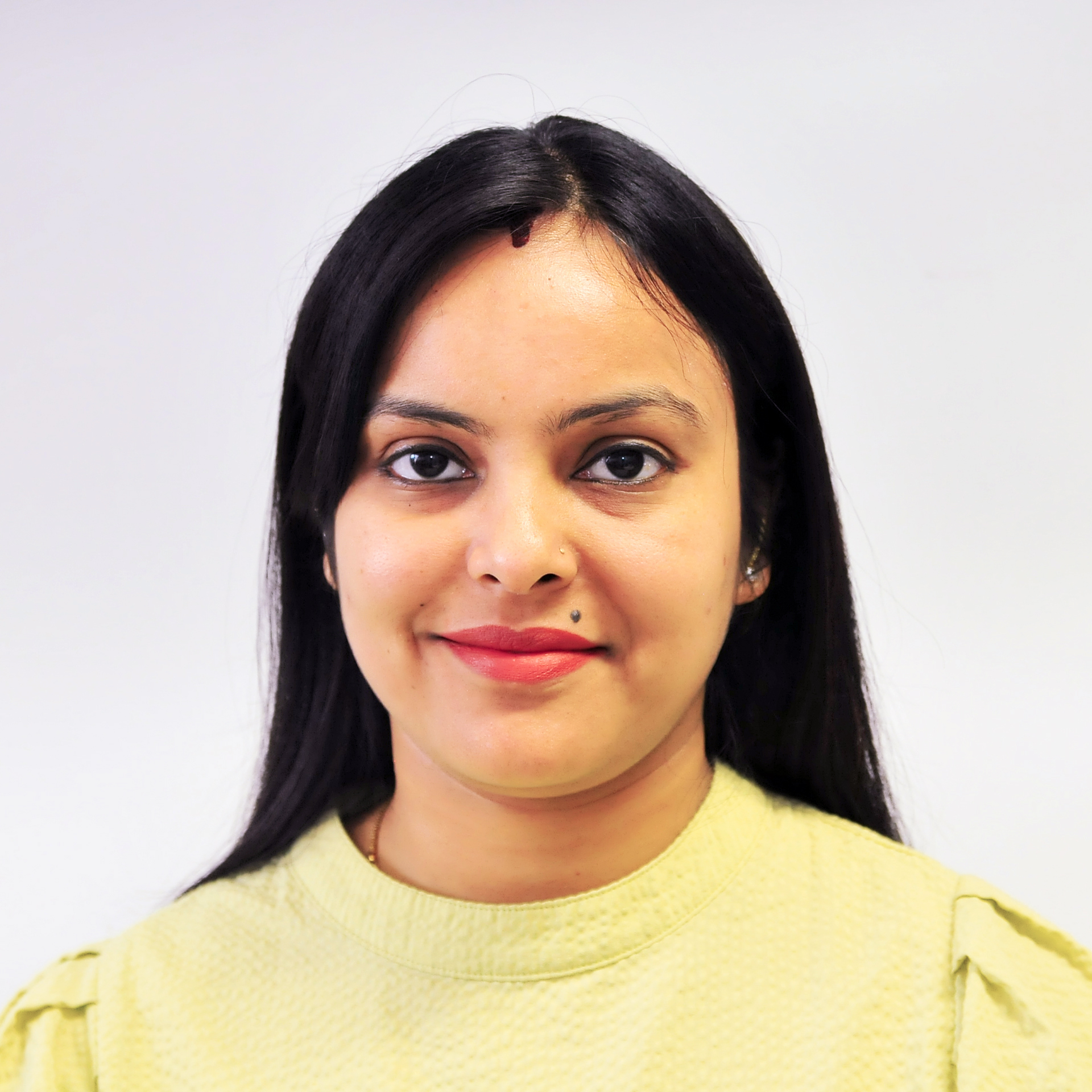}}]{Deepika Saxena} is working as an Associate Professor in the Division of Information Systems at the University of Aizu, Japan. She received her Ph.D. degree in Computer Science from the National Institute of Technology, Kurukshetra, India, and completed her Post Doctorate from the Department of Computer Science at Goethe University, Frankfurt, Germany. She has received several prestigious awards, including IEEE TCSC Early Career Researcher Award 2024, IEEE TCSC Outstanding Ph.D. Dissertation Award 2023, EUROSIM Best Ph.D. Thesis Award 2023, and IEEE Computer Society Best Paper Award 2022. She is also a recipient of the JSPS KAKENHI Early Career Young Scientist Research Grant FY2024. Her research focuses on Neural Networks, Evolutionary Algorithms, Cloud Computing, Digital Twins, Quantum Machine Learning, Cyber Threat Intelligence, and Vehicle Traffic Management.  
\end{IEEEbiography}

\begin{IEEEbiography}[{\includegraphics[width=0.7\linewidth]{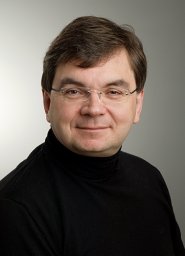}}]{Volker Lindenstruth}is working as a Professor in the Department of Computer Science and Department of Physics, Goethe University, Frankfurt, Germany. Professor Volker Lindenstruth studied physics at the TU Darmstadt and received his diploma in 1989. From 1989 to 1993 he obtained his doctorate at the GSI in Darmstadt in physics and then went on to work as a postdoc for computer science for two years as Feodor v. Lynen Fellow at LBNL, USA.  He has been the head of the ALICE HLT project (BMBF / CERN MoU) at the LHC of CERN and from 2006 to 2007 also CERN Associate. In 2005, he founded Certon Systems in Heidelberg. At FIAS he held the position of Fellow since 2007 and became a Senior Fellow soon. Furthermore, the group of High-Performance Computer Architecture of the Goethe University has been in his care since 2009. His research area includes Computer Sciences and AI Systems, Energy, Big Data, High Performance Computing, Nuclear Physics.  \end{IEEEbiography}


\end{document}